\documentclass[%
 reprint,longbibliography,preprintnumbers,
nofootinbib,
 amsmath,amssymb,
 aps,
pre,
]{revtex4-2}
\pdfoutput=1
\usepackage{graphicx}
\usepackage[utf8]{inputenc}
\usepackage{flushend}
\usepackage{dcolumn}
\usepackage{bm}
\usepackage{balance}
\usepackage{placeins}

\usepackage{comment}

\usepackage{float}

\usepackage{xcolor} 
\usepackage[normalem]{ulem} %

\usepackage[normalem]{ulem}

\usepackage{graphicx,tikz}
\usetikzlibrary{calc,arrows.meta,fit,backgrounds}

\usepackage{verbatim}
\usepackage{color,ulem}
\usepackage[english]{babel}
\newcommand{\liu}[1]{{\textcolor{blue}{#1}}}

\newcommand{\Ying}[1]{{\color{brown}{#1}}}

\def\be{\begin{equation}}
\def\ee{\label{eq:prl-extra-1}
\end{equation}}
\def\bea{\begin{eqnarray}}
\def\eea{\end{eqnarray}}

\usepackage[utf8]{inputenc}

\input Starburst.fd
\newcommand*\initfamily{\usefont{U}{Starburst}{xl}{n}}\initfamily

\makeatletter 
    
\renewcommand\onecolumngrid{
\do@columngrid{one}{\@ne}%
\def\set@footnotewidth{\onecolumngrid}
\def\footnoterule{\kern-6pt\hrule width 1.5in\kern6pt}%
}

\renewcommand\twocolumngrid{
        \def\footnoterule{
        \dimen@\skip\footins\divide\dimen@\thr@@
        \kern-\dimen@\hrule width.5in\kern\dimen@}
        \do@columngrid{mlt}{\tw@}
}%

\makeatother

\newcommand{\beq}{\begin{eqnarray}}
\newcommand{\eeq}{\end{eqnarray}}
\usepackage{amsmath}
\usepackage{tikz}
\usetikzlibrary{decorations.pathmorphing}
\usetikzlibrary{shapes.misc}
\tikzset{cross/.style={cross out, draw=black, minimum size=8*(#1-\pgflinewidth), inner sep=0pt, outer sep=0pt},
cross/.default={1pt}}
\usetikzlibrary{patterns,math}
\usepackage[colorlinks = true,
            linkcolor = blue,
            urlcolor  = blue,
            citecolor =red,
            anchorcolor = blue,
            hypertexnames=true]{hyperref}
            
\allowdisplaybreaks[4]

\begin{document}

\title{Multipartite dynamics beyond bipartite entanglement equilibration
}

\author{Yan Liu$^{1}$}
\email{yanliu@buaa.edu.cn}

\author{Ya-Wen Sun$^{2}$}
\email{yawen.sun@ucas.ac.cn}

\author{Ren-Jie Ying$^{3}$}
\email{2400018511@stu.pku.edu.cn}

\author{Jun-Yuan Zeng$^{2}$}
\email{zengjunyuan2025@mails.ucas.ac.cn}

\affiliation{
\vspace{0.25cm}
$^1$Department of Space Science,  
and Peng Huanwu Collaborative Center for Research and Education, Beihang University, Beijing 100191, China}
\affiliation{$^2$School of Physical Sciences, University of Chinese Academy of Sciences, Beijing 100190, China}
\affiliation{$^3$Peking University, Beijing 100871, China}

\begin{abstract}
{Entanglement equilibration need not mark the end of quantum-information dynamics in quantum thermalization. We show that the complete Schmidt spectra of a fixed tripartition can become stationary while its multipartite correlation structure, distinguished by local-unitary invariants, continues to evolve. This separation occurs exactly within a finite window in self-dual kicked-Ising dynamics and asymptotically in Haar-random circuits. We further demonstrate an independently tunable timescale for multipartite dynamics using quantum-error-correcting codes. When coupled to a locally scrambling bath, ordinary entanglement approaches equilibrium through rapid local scrambling, while weak logical--bath coupling drives a slower reorganization of multipartite correlations. These results reveal a deeper distinction than the ordering of equilibration times: bipartite entanglement and multipartite correlation structure can probe distinct physical channels of quantum-information dynamics, rather than merely different stages of the same relaxation process.}
\end{abstract}

\maketitle

\section{Introduction}
\label{sec:prl-intro}

Thermalization in an isolated quantum many-body system does not bring microscopic motion to an end; rather, it makes that motion increasingly invisible to the observables through which equilibrium is judged. Following a quench, entanglement spreads through the system and subsystem entropies eventually saturate \cite{Calabrese:2009qy, Abajo-Arrastia:2010ajo, KimHuse2013, DAlessio2016}. Yet an entanglement plateau characterizes how a region is entangled with the rest, not how quantum correlations are organized jointly across different regions. More strikingly, individual regions may appear fully equilibrated while their multipartite structure keeps evolving.

Recent holographic studies and long-lived Markov-gap dynamics in fermionic systems have revealed multipartite
dynamics beyond ordinary entanglement-entropy plateaus
\cite{AnegawaMiyataSuzukiTamaoka2026,
{BalasubramanianJiangRoss2026},
{JuSunZhaoZhou2026Dynamics}, 
{BasuWenXu2026AdSBCFT},ZhouZhangZhang2026}. Yet the saturation of entanglement entropies does
not imply that the complete Schmidt spectra have
stopped evolving. We therefore pose a much stronger
question: can multipartite correlation structure
continue to evolve even when all three Schmidt spectra,
or even the set of one-party reduced density matrices,
remain stationary? More intriguingly, can this remaining dynamics be governed by a distinct physical channel
with its own independently tunable timescale?

Here we establish two distinct types of dynamical separation in explicit unitary many-body models. In {\bf Type I}, we show that multipartite
correlation structure can continue to evolve after all
three complete Schmidt spectra have become stationary,
exactly in self-dual kicked-Ising dynamics and
asymptotically in Haar-random circuits. In {\bf Type II}, we go further by identifying an independently tunable timescale for the remaining multipartite dynamics. For a quantum-error-correcting-code system coupled to a locally scrambling bath, ordinary entanglement approaches equilibrium through rapid local scrambling, while multipartite correlations continue to reorganize through the weaker logical--bath coupling with an independent slower timescale.

These results reveal a deeper distinction than the ordering of equilibration times. Bipartite and multipartite  structures need not simply probe different stages of the same relaxation process; they can reflect distinct physical channels of quantum information dynamics. The central question is therefore not which entanglement measure equilibrates last, but which physical processes govern the evolution of different correlation structures.

\vspace{-0.3cm}
\section{Bipartite Spectra and Multipartite Dynamics}
\label{sec:prl-framework}

\noindent For a pure state $|\psi\rangle$ of a tripartite system
$ABC$, the complete Schmidt spectra across the three
cuts $A|BC$, $B|CA$, and $C|AB$ are given by
$\{\operatorname{Spec}\rho_A,\operatorname{Spec}\rho_B,
\operatorname{Spec}\rho_C\}$, where
$\rho_X=\operatorname{Tr}_{\bar X}|\psi\rangle\langle\psi|$.
These spectra fully characterize bipartite entanglement
across the three cuts, but do not determine how
correlations are jointly organized among $A$, $B$, and $C$.
This opens a striking dynamical possibility:
multipartite correlation structure can continue to evolve even when all three complete Schmidt spectra are stationary. More remarkably, such evolution can persist even when the full reduced density matrices $\rho_A$, $\rho_B$, and $\rho_C$ themselves remain fixed.

We characterize the multipartite correlation structure
of a state by how quantum correlations are organized
among $A$, $B$, and $C$, independently of local
unitary transformations within each party.
States related by $U_A\otimes U_B\otimes U_C$ therefore
have the same multipartite structure, naturally
identified with their local-unitary (LU) equivalence class.
A systematic family of invariants characterizing
these structures is 
\begin{equation}
\begin{aligned}
\mathcal I_{\boldsymbol\pi}^{(r)}
=\operatorname{Tr}\bigl[\rho^{\otimes r}\mathcal P_{\boldsymbol\pi}\bigr],\quad
\mathcal P_{\boldsymbol\pi}
=P_A(\pi_A)P_B(\pi_B)P_C(\pi_C),
\end{aligned}
\label{eq:prl-display-1}
\end{equation}
where $\rho=|\psi\rangle\langle\psi|$,
$r$ is the number of copies, and
$P_X(\pi_X)$ permutes the copies of party $X$
according to $\pi_X\in S_r$.
Taken together over all orders and permutations, these invariants distinguish multipartite entanglement structures up to local unitaries in finite dimensions \cite{GrasslRottelerBeth1998,Vrana2011,Vrana2011Generating,GaddeKrishnaSharma2023Classification}, although individual invariants need not be measures of genuine multipartite entanglement. Measurement protocols based on multiple copies have also been developed for such LU invariants \cite{LeiferLindenWinter2004}.

As a concrete diagnostic, we mainly use the four-replica genuine multi-entropy
\cite{GaddeKrishnaSharma2022,IizukaNishida2025,
PeningtonWalterWitteveen2023},
\begin{equation}
\begin{aligned}
G\equiv GM_2^{(3)}
&=-\frac12\log Z_2^{(3)}
-\frac12\sum_{X=A,B,C}S_2(X),\\
Z_2^{(3)}
&=\mathcal I_{((13)(24),(12)(34),e)}^{(4)},
\end{aligned}
\label{eq:prl-display-2}
\end{equation}
where $e$ is the identity permutation and
$S_n(X)=(1-n)^{-1}\log\operatorname{Tr}\rho_X^n$
is the R\'enyi entropy, with
$S_1(X)=-\operatorname{Tr}(\rho_X\log\rho_X)$.
Multi-entropy, replica-based invariants, and related multipartite entanglement diagnostics have been studied in holography and many-body systems
\cite{IizukaLinNishida2025,GaddeHarperKrishna2025,
HarperTakayanagiTsuda2024,HarperMollabashiTakayanagiTasuki2025, Jiang:2025iet, AnegawaSuzukiTamaoka2026, JuLiuSunXuZhao2026IR, 
Chen:2026xtx, 
IizukaMiyata2026Junction, JuSunZhao2026Sperner,
IizukaMiyata2026Confining,
Chen:2026hcf, HuLinNechita2026,
JuSunZhaoZhou2026Dynamics, Chen:2026rew}.
Their sensitivity depends on the replica index: for example,
$GM_2^{(3)}$ vanishes for pure bosonic Gaussian states
\cite{CamargoNishida2026}.

Even fixing all three one-party reduced density matrices
does not determine the multipartite correlation structure.
For instance, consider the three-qutrit GHZ
state $|\Phi\rangle$
and the fully antisymmetric state
$|\Omega\rangle$.
Their superpositions
$|\psi_\theta\rangle=\cos\theta|\Phi\rangle+
\sin\theta|\Omega\rangle$ satisfy (see App.~\ref{sec:sm_three-qutrit})
\begin{equation}
\begin{aligned}
\rho_A(\theta)=\rho_B(\theta)=\rho_C(\theta)
&=\frac{\mathbb I_3}{3},\\
G(0)=0,\qquad G(\pi/2)
&=\frac12\log2.
\end{aligned}
\label{eq:prl-display-3}
\end{equation}
Thus, states with identical  reduced density
matrices can have distinct multipartite LU structures.
This continuous family illustrates how multipartite
correlations can change with unchanged bipartite Schmidt spectra.

To understand how this structure can be explored dynamically,
consider $H=H_A+H_B+H_C+H_\partial$, where $H_\partial$
contains interparty interactions, and define
$H_\partial^{[r]}=\sum_{j=1}^r H_\partial^{(j)}$.
The Schmidt spectra are encoded in their moments
$M_{n,X}=\operatorname{Tr}\rho_X^n
=\operatorname{Tr}[\rho^{\otimes n}P_X(\gamma_n)]$,
while $Z_2^{(3)}$ probes multipartite correlations.
Their time evolution is
\begin{equation}
\begin{aligned}
\dot M_{n,X} &=i\bigl\langle[H_\partial^{[n]},P_X(\gamma_n)]\bigr\rangle_n, \quad 
\dot Z_2^{(3)}=i\bigl\langle[H_\partial^{[4]},\mathcal P_G]\bigr\rangle_4, \\
\dot G &=-\frac{\dot Z_2^{(3)}}{2Z_2^{(3)}} +\frac12\sum_{X=A,B,C} \frac{\dot M_{2,X}}{M_{2,X}},
\end{aligned}
\label{eq:prl-display-4}
\end{equation}
where $\gamma_n=(12\cdots n)$,
$\mathcal P_G=P_A((13)(24))P_B((12)(34))$,
and $\langle O\rangle_r=\operatorname{Tr}(\rho^{\otimes r}O)$.
The intraparty terms drop out because local unitaries
preserve both Schmidt spectra and multipartite LU structure.
In contrast, the same interparty interactions can affect their distinct replica operators differently, allowing multipartite correlations to continue evolving even when all Schmidt spectra remain stationary. We demonstrate this separation in concrete quantum many-body models in the next section.

\vspace{-0.2cm}
\section{Multipartite Information Dynamics in Many-Body Models}
\label{sec:prl-models}

\noindent Previous studies have shown that multipartite correlations can continue to evolve after entanglement entropies saturate \cite{AnegawaMiyataSuzukiTamaoka2026, {JuSunZhaoZhou2026Dynamics}}. We address a far more stringent question: can multipartite structure still evolve when all three Schmidt spectra of a fixed tripartition, or even the complete one-party reduced density matrices, are stationary? {\bf Type I} progresses from distinct stages of equilibration in a tilted-field Ising chain to multipartite evolution at stationary Schmidt spectra, realized exactly in self-dual kicked-Ising dynamics and asymptotically in Haar-random circuits. {\bf Type II} goes further: the remaining multipartite dynamics can acquire an independently tunable timescale and, in a locally scrambling bath, be driven by a distinct physical channel.

\vspace{-0.3cm}
\subsection{Type I: From Entanglement Plateaus to Spectrum-Invisible Dynamics}


\vspace{-0.3cm}
\noindent 
\emph{Different stages of equilibration: tilted-field Ising chain.—} A nonintegrable spin chain provides a first view of how different entanglement quantities settle during the same unitary evolution. We evolve equatorial product states under the open-chain Hamiltonian (see App.~\ref{sec:sm_tilted})
\begin{equation}
    H_{\mathrm{TI}}=J\sum_{j=1}^{N-1}Z_jZ_{j+1}+\sum_{j=1}^{N}(h_x X_j+h_z Z_j)+h_b Z_1.
    \label{eq:prl-display-5}
\end{equation}
We set $\hbar=1$ and $(J,h_x,h_z,h_b)=(1,1.05,0.5,0.37)$, with $J$ setting the energy unit.
The mean adjacent-gap ratios for the central $80\%$ of the spectra are $\langle r\rangle=(0.5260,0.5299,0.5291)$ for $N=(8,10,12)$, respectively, close to the GOE benchmark ($\simeq0.5307$), supporting the model's nonintegrable character.

{
For three equal contiguous regions and five equatorial
product initial states with identical mean energy, a common
finite-window $10\%$ equilibration criterion gives, at
$N=18$, mean plateau times
$t_{S_1}\simeq6.35$, $t_G\simeq9.05$,
$t_{\mathrm{spec}}\simeq9.85$, and
$t_{S_\infty}\simeq10.70$.
Here $t_{\mathrm{spec}}$ is determined from the distance
between the complete Schmidt spectra and their late-time
references, whereas $S_\infty=-\log\lambda_{\max}$ probes
the largest Schmidt eigenvalue. Although derived from the
same spectra, the two quantities can reach their operational
plateaus at different times.
}

These results reveal distinct stages of equilibration:
$G$ settles later than the ordinary von Neumann entropy,
but earlier than the complete Schmidt spectra. However,
$G$ probes only one aspect of multipartite structure.
Further calculations at $N=9$ show that the higher-replica
$GM_3^{(3)}$ first enters its nominal $10\%$ plateau
later on average than $GM_2^{(3)}$ (see App.~\ref{sec:sm_tilted}). Higher-order multipartite
invariants can thus reveal later stages of multipartite correlation
dynamics, suggesting that the multipartite structure
as a whole may continue to evolve even after lower-order invariants or the complete Schmidt spectra have settled.

\vspace{0.2cm}
\noindent 
{\it Exact multipartite dynamics after local equilibration: self-dual kicked Ising.—} We next realize a stronger separation exactly: the complete reduced states of all three parties become stationary, while their multipartite structure continues to evolve. Consider the periodic self-dual kicked-Ising (SDKI) Floquet unitary~\cite{BertiniKosProsen2019Entanglement,BertiniKosProsen2019DualUnitary}
\begin{equation}
\begin{aligned}
U_F&=U_KU_I,\qquad
U_K=\exp\left(-\frac{i\pi}{4}\sum_{j=1}^{N} X_j\right),\\
U_I&=\exp\left[-i\left(
\frac{\pi}{4}\sum_{j=1}^{N}Z_jZ_{j+1}
+h\sum_{j=1}^N Z_j
\right)\right].
\end{aligned}
\label{eq:prl-display-6}
\end{equation}
{Starting from $|\psi_0\rangle=|0\rangle^{\otimes N}$, we evolve the system according to $|\psi_t\rangle=U_F^t|\psi_0\rangle$ at integer times $t$, with Floquet period $T=1$. Each time step corresponds to one complete Floquet period, in which the diagonal Ising and longitudinal-field segment acts first, followed by the transverse kick. Dividing the ring into three equal contiguous regions of $\ell=N/3$ sites, dual unitarity yields an exact interval of maximally mixed one-party states,}
\begin{equation}
\rho_A(t)=\rho_B(t)=\rho_C(t)
=2^{-\ell}\mathbb I,
~~~~
1+\left\lceil\frac{\ell}{2}\right\rceil
\leq t\leq\ell+1.
\label{eq:prl-display-7}
\end{equation}
Every Schmidt spectrum of the three bipartitions is therefore exactly stationary within this finite local-equilibration window.

\begin{figure}[h!]
\centering  \includegraphics[width=0.80\columnwidth]
{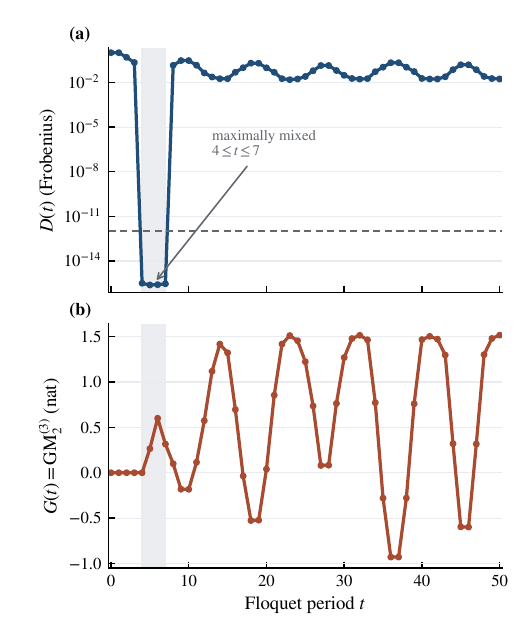}
\caption{\label{fig:typeI_sdki}
{\small Multipartite dynamics during an exact marginal plateau in the periodic SDKI chain ($N=18$, $\ell=6$, $h=0.15$). (a) Maximum Frobenius deviation $D(t)=\max_{X=A,B,C} \|\rho_X(t)-\mathbb I_{64}/64\|_{\mathrm F}$; dashed line marks the $10^{-12}$ tolerance. (b) Genuine tripartite multi-entropy. 
}
}
\end{figure}

For $N=18$, $\ell=6$, and $h=0.15$, Fig.~\ref{fig:typeI_sdki}
illustrates the exact marginal plateau in Eq.~\eqref{eq:prl-display-7}
(see App.~\ref{sec:sm_sdki}). All three reduced density matrices remain maximally mixed to numerical precision at $t=4,5,6,7$, with $\max_{X,t}\|\rho_X(t)-\mathbb I_X/2^{|X|}\|_{\mathrm F} =3.23\times10^{-16}$ over this window. Nevertheless, the multipartite invariant varies substantially,
\begin{equation}
\begin{aligned}
&G(4)\simeq0,\qquad G(5)\simeq0.265437,\\
&G(6)\simeq0.600253,\quad G(7)\simeq0.315883.
\end{aligned}
\label{eq:sdki-n18-G}
\end{equation}
Thus, although each party remains individually maximally mixed, the LU invariant multipartite structures  shared among the three parties continue to reorganize. 

\vspace{0.2cm}
\noindent
\emph{Multipartite growth under Haar scrambling.---}
The same separation also emerges in Haar-random dynamics: multipartite correlations continue to evolve even as the bipartite spectra approach stationary values
\cite{NahumRuhmanVijayHaah2017,ZhouNahum2019}.
Consider a periodic brickwork circuit $U(t)$ of independent Haar-random $U(d^2)$ gates acting on $N$ qudits, where $t$
counts individual circuit layers. Its normalized Choi state $|U(t)\rangle=d^{-N/2}
\sum_{ij}U_{ji}(t)|i\rangle_L|j\rangle_R$ is partitioned into aligned intervals $A\subset L$ and
$B\subset R$, each of length $\ell$, and their complement
$C$ containing $2m$ qudits, with $m=N-\ell$.
Unitarity fixes $\rho_A=\rho_B=\mathbb I/d^\ell$
for every circuit realization. The question is whether
$\rho_C$ can also approach a stationary state while
multipartite correlations continue to change
\cite{HosurQiRobertsYoshida2016}.

To resolve these two dynamical aspects, we evaluate
the ordinary second R\'enyi entropy and the four-replica
correlation entering genuine multi-entropy. Define
$F_2=-\log\mathbb E\operatorname{Tr}\rho_C^2$ and
$F_4=-\log\mathbb E Z_2^{(3)}$, which combine into the
annealed multi-entropy
$G_{\mathrm{ann}}=(F_4-F_2)/2-\ell\log d$.
For $N=18$, $\ell=12$, and $m=6$, exact Haar averaging
gives, in the large-$d$ limit at fixed circuit depths,
\begin{equation}
\begin{aligned}
F_4(8)&=40\log d+O(d^{-2}),\\
F_4(10)&=44\log d+O(d^{-2}).
\end{aligned}
\label{eq:prl-display-9}
\end{equation}

To show the complete Schmidt spectra have
settled, we measure the deviation of $\rho_C$ from maximal mixing
\begin{equation}
\Delta_C=d^{2m}\mathbb E\operatorname{Tr}\rho_C^2-1,
\quad
\mathbb E\left\|
\rho_C-\frac{\mathbb I}{d^{2m}}
\right\|_1
\leq\sqrt{\Delta_C}.
\label{eq:prl-display-10}
\end{equation}
The exact Haar-averaged calculation yields
\begin{equation}
\begin{aligned}
\Delta_C(8)
&=4836\,d^{-4}+O(d^{-6}),\\
\Delta_C(10)
&=61479\,d^{-8}+O(d^{-10}),\\
G_{\mathrm{ann}}(10)-G_{\mathrm{ann}}(8)
&=2\log d+O(d^{-2}).
\end{aligned}
\label{eq:prl-display-11}
\end{equation}
Thus all reduced density matrices approach maximal mixing in
average trace norm as $d\to\infty$ at fixed circuit size and depth,
while an annealed multipartite response of size $2\log d$
persists. Here $G_{\rm ann}$ places logarithms after ensemble averaging.  
The full-marginal
decoupling and signed four-replica calculations are in
Apps.~\ref{sec:sm_haar_typeI} and~\ref{sec:sm_haar_noncancellation}.

\vspace{-0.3cm}
\subsection{Type II: Protected Information and a Separate Dynamical Clock}

\noindent Type I shows delayed multipartite equilibration, yet without an independent new relaxation timescale. Can the residual multipartite dynamics be governed by its own physical channel and timescale? We first examine charge diffusion with hydrodynamic slow dynamics, yet find that the slowest diffusive mode does not control the dominant equilibration of multipartite dynamics. We then construct a quantum-error-correcting circuit in which multi-entropy evolves and equilibrates on a tunable slower timescale, independent of the bipartite dynamics.

\vspace{0.2cm}
\noindent \emph{Hydrodynamic slow modes.—} Consider an isolated chain of \(N\) qubits with open boundary conditions, evolved by a \(U(1)\)-conserving brickwork circuit. Each circuit period consists of two staggered layers of local gates \(e^{i\phi_0}\oplus V_2\oplus e^{i\phi_2}\), where \(V_2\) is Haar distributed in \(U(2)\)~\cite{KhemaniVishwanathHuse2018,RakovszkyPollmannKeyserlingk2019}. Averaging over the random gates yields an exact diffusive evolution of the charge density, with the slowest nonuniform mode decaying at $\gamma_Q=-2\log\cos(\pi/N)
\simeq\frac{\pi^2}{N^2}.$

We define operational entanglement saturation times by the first $90\%$ crossings toward late-time reference values averaged over $80\leq t\leq100$. For $N=18$, $S_1$, $S_2$, and $G$ reach these thresholds at $t_{S_1}=20$, $t_{S_2}=28$, and $t_G=36$, respectively. The independently identified slowest conserved-charge mode has an exponential decay time $\tau_Q=32.661$, corresponding to a $90\%$ amplitude-decay time $t_{Q,90}=\tau_Q\ln10\simeq75.2$. Thus, the dominant entanglement growth is largely complete while the diffusive charge mode continues to relax. 
Across the five sizes, finite-size fits yield $t_{S_1}\propto N^{1.067}$, $t_G\propto N^{1.159}$, and $\tau_Q\propto N^{2.039}$
(See App.~\ref{sec:sm_typeII_hydro} for details).

Despite the presence of a diffusive charge mode with a characteristic slow \(O(N^2)\) timescale, we find no clear evidence that it controls the dominant growth times of $S_1$ or $G$, which scale approximately linearly with system size. Diffusion can nevertheless govern entanglement dynamics: e.g., the sub-ballistic growth of higher Rényi entropies in \cite{RakovszkyPollmannKeyserlingk2019}. An intriguing direction is thus to construct many-body dynamics where multipartite correlations couple selectively to hydrodynamic modes, acquiring a distinct diffusion-controlled equilibration channel.

\begin{figure*}[t]
    \centering
    \includegraphics[width=0.775\textwidth]
{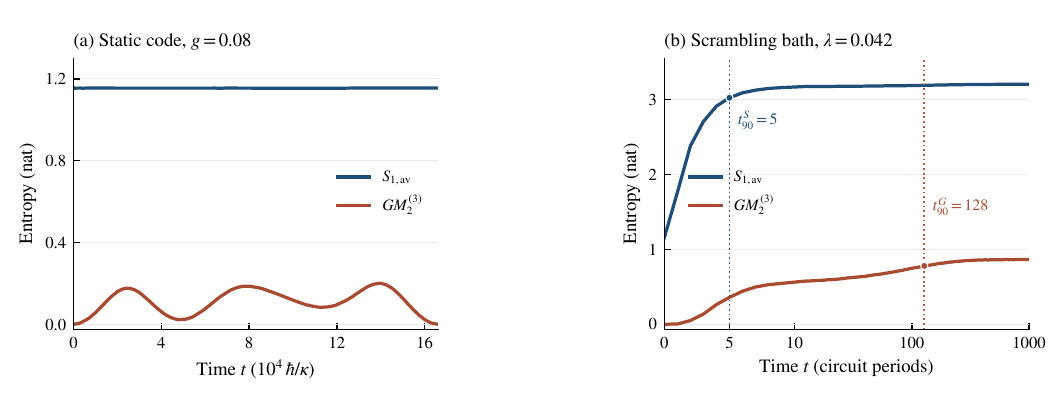} 
   \vspace{-0.4cm}
    \caption{\small
    Ordinary entropy and multipartite dynamics in
    two code-based models. Dark blue and reddish brown curves denote $S_{1,\mathrm{av}}=\frac13\sum_{X=A,B,C}S_1(X)$ 
    and $G=GM_2^{(3)}$, respectively.
    (a) Five-qubit static code
    ($g=0.08$, $\kappa=1$, with $2|2|1$ partition). 
    (b) The code coupled to a nine-qubit locally
    scrambling bath ($\lambda=0.042$), averaged
    over 48 independent circuit realizations.
    The time axis is linear for $t\leq10$
    and logarithmic thereafter.}
 \label{fig:code_bath_compact}
\end{figure*}

\vspace{0.2cm}
\noindent {\it Protected logical dynamics in a five-qubit static code.—}
We next demonstrate a parametrically separated multipartite clock using the \([[5,1,3]]\) quantum error-correcting code (QEC)~\cite{LaflammeMiquelPazZurek1996,BennettDiVincenzoSmolinWootters1996,KnillLaflamme1997}. The code corrects the erasure of any two physical qubits, so that logical information cannot be recovered by any individual party in the tripartition \(A=(0,1)\), \(B=(2,3)\), and \(C=(4)\), but can be recovered from suitable joint parties.

We consider the Hamiltonian
\begin{equation}
H(g)=-\kappa\sum_{a=1}^{4}S_a
+g\sum_{i=0}^{4}\sum_{\alpha=x,y,z}
h_i^\alpha\sigma_i^\alpha,
\label{eq:prl-display-13}
\end{equation}
where \(S_a=(XZZXI,IXZZX,XIXZZ,ZXIXZ)\) are the stabilizer generators, and {$g$ controls the strength of the local perturbation}. We set \(\kappa=1\), \(\hbar=1\), and use a fixed normalized set of local fields \(h_i^\alpha\) specified in App.~\ref{sec:sm_typeII_static_code}. The unperturbed $(g=0)$ ground space encodes one logical qubit and is separated from the excited syndrome sectors by \(\Delta=2\kappa\).

Every state strictly within the ideal code space satisfies
\begin{equation}
\rho_A=\frac{\mathbb I_4}{4},\qquad
\rho_B=\frac{\mathbb I_4}{4},\qquad
\rho_C=\frac{\mathbb I_2}{2}.
\label{eq:prl-display-14}
\end{equation}
Thus logical rotations strictly within the ideal code space leave all three single-party reduced density matrices, and hence their entropies, unchanged. Under the full perturbed Hamiltonian, however, virtual admixture of excited syndrome sectors generally displaces the evolving state from this exact fixed-marginal set. In the weak-coupling regime $g/\Delta\ll1$, the reduced density matrices acquire corrections \(O(g/\Delta)\), while their entropy deviations are suppressed to \(O(g^2/\Delta^2)\).

The distance-three code structure forbids nontrivial logical dynamics at first and second order in the one-body perturbation. The leading effective logical Hamiltonian arises at third order,
\begin{equation}
\begin{split}
H_{\mathrm{eff}}^{(L)}
=(E_0+g^2c_2)\mathbb I_2
+g^3\boldsymbol h_3\cdot\boldsymbol\sigma_L
+O(g^4/\Delta^3),
\label{eq:prl-display-15}
\end{split}
\end{equation}
where \(E_0=-2\Delta\){, $c_2=O(\Delta^{-1})$ is the second-order energy-shift coefficient common to both logical states, and $\boldsymbol h_3=O(\Delta^{-2})$}. Consequently, fast syndrome dressing occurs on \(t_{\mathrm{local}}\sim\Delta^{-1}\), whereas coherent logical rotation occurs on the parametrically longer scale \(\tau_L\sim\Delta^2/g^3\). The two characteristic times originate from different effective processes within the same static Hamiltonian.

Starting from the bare logical state $|\overline 0\rangle$, exact evolution exhibits substantial oscillations of \(G=\mathrm{GM}^{(3)}_2\) while the ordinary entropies show only small dressing-induced fluctuations [Fig.~\ref{fig:code_bath_compact}(a)]. {The code-restricted four-replica contraction makes $G$ explicitly sensitive to the orientation of the logical Bloch vector, so its coherent precession at frequency $\Delta E_L$ induces GM oscillations on a timescale $t_{1/2}^{G}\sim\Delta E_L^{-1}$ (see App.~\ref{sec:sm_typeII_static_code}).} For \(0.035\leq g\leq0.15\), independent numerical fits yield
\begin{equation}
\begin{aligned}
\Delta E_L&\simeq0.0738770\,g^{3.00010},~~~
t_{1/2}^{G}\simeq7.06608\,g^{-3.00427},
\end{aligned}
\label{eq:prl-display-16}
\end{equation}
where \(t_{1/2}^{G}\) is the first half-rise time of a coherent GM excursion. The mean entropy drift scales approximately as \(\Delta S_{\mathrm{av}}\propto g^{1.9544}\), consistent with the $O(g^2/\Delta^2)$ expectation, whereas the GM excursion $\Delta G$ remains \(O(1)\), approximately \(0.2\). At \(g=0.08\), for example, \(\Delta G=0.200482\) while \(\Delta S_{\mathrm{av}}=0.00130996\). 

These results reveal a parametrically slow channel of coherent multipartite dynamics, accompanied by only small, microscopic-scale changes in ordinary entanglement. In this bath-free model, \(G(t)\) undergoes recurrent oscillations rather than relaxation, so \(t_{1/2}^{G}\) characterizes the timescale of logical motion, not thermalization. We next ask whether a similar separation can emerge in the equilibration dynamics of a locally scrambling many-body system.

\vspace{0.2cm}
\noindent \emph{The same code coupled to a locally scrambling bath.—}
We now retain the five-qubit setup but couple it to a nine-qubit bath undergoing local random-unitary scrambling. The code is partitioned as \(2|2|1\) and the bath into three consecutive blocks of three qubits, together forming the pure tripartition \(ABC\). The initial state is $|\overline 0\rangle$ tensored with independently sampled single-qubit pure bath states. Each circuit period consists of two staggered layers of independently sampled nearest-neighbor Haar $U(4)$ gates, followed by a code-preserving interaction,
\begin{align}
\begin{split}
U_t&=U_c(\lambda)U_{\rm odd}(t)U_{\rm even}(t),
\label{eq:prl-display-17}\\
U_c(\lambda)&=\exp\left[-i\lambda\left(
\overline X\otimes B_L+
\overline Z\otimes B_R
\right)\right],
\end{split}
\end{align}
{where $\lambda$ is the dimensionless logical--bath coupling strength, $\overline X=X^{\otimes5}$ and $\overline Z=Z^{\otimes5}$ are logical Pauli operators, and $B_L=Y_0X_1$ and $B_R=X_7Y_8$ act on the bath.}

Unlike the physically perturbed static code, this interaction preserves the ideal code space exactly. The erasure properties ensure that each code share remains maximally mixed and uncorrelated with its corresponding bath share, $\rho_X(t)=\frac{\mathbb I_{c_X}}{d_{c_X}}\otimes\rho_{b_X}(t)${, $X=A,B,C$}. Consequently, the ordinary single-party entanglement entropies satisfy $S_n(X,t)=\ln d_{c_X}+S_n(\rho_{b_X}(t))$. Thus, while the code shares remain exactly maximally mixed, the entanglement entropies of the complete parties can evolve through bath scrambling. The genuine multipartite invariant $G$ can additionally probe the slower evolution of encoded logical correlations.

{For each $\lambda$, we average over 48 independently sampled random circuits, each paired with a $\lambda=0$ reference using identical initial bath states and Haar-random gates.} 
For \(\lambda=(0.025,0.032,0.042,0.055,0.072)\), the ordinary mean von Neumann entropy reaches its operational \(90\%\) reference level after approximately $t^{S_1}_{90}=5$ periods for every coupling. By contrast, the corresponding GM times are \(t_{90}^{G}=(360,224,128,80,44)\). The separation at $\lambda=0.042$ is illustrated in Fig.~\ref{fig:code_bath_compact}(b). To isolate the logical-contact contribution, we subtract the \(\lambda=0\) trajectory with identical random gates; the resulting paired response has \(t_{90}^{\Delta G}=(616,368,224,124,72)\), with an effective scaling \(t_{90}^{\Delta G}\propto\lambda^{-2.025}\).

The origin of the slow clock is independently visible in the \emph{exactly averaged} logical channel, whose polarization decays with
\begin{equation}
\Gamma_L=-2\ln|\cos(\sqrt2\lambda)|
=2\lambda^2+O(\lambda^4).
\label{eq:prl-display-19}
\end{equation} 
This is an exact decay rate for the ensemble-averaged logical polarization. The paired GM crossing times  exhibit the same weak-contact $\lambda^{-2}$ scale (see App.~\ref{sec:sm_typeII_local_bath} for details), while the dominant ordinary-entropy growth is rapid. 
A complementary collision model realizing tunable multipartite relaxation at strictly fixed complete-party marginals could be found  
in App.~\ref{sec:sm_typeII_collision}.

In the locally scrambling bath, ordinary entanglement approaches equilibrium through rapid local scrambling, while weak coupling to encoded logical information continues to reorganize multipartite correlations on a much longer timescale. This separation is therefore more than a difference in equilibration times: it reflects the distinct physical processes governing the evolution of different correlation structures. More broadly, bipartite entanglement and multipartite information need not simply provide different measures of the same relaxation process; they may probe different dynamical channels through which quantum information spreads, is stored, and is reorganized.

\vspace{0.8cm}

\vspace{-0.5cm}
\section*{acknowledgments} 
We thank Ling-Yan Hung for useful discussions.
This work was supported by the National Natural Science Foundation of China Grants No. 12375041, 12575046 
and 12575068. {
GPT-5.6 and 6 Sol (OpenAI) was used to assist with numerical computations, code development, and manuscript preparation. All scientific ideas, physical interpretations, and conclusions are those of the authors, who independently verified the reported results and take full responsibility for the manuscript.
}

\bibliographystyle{apsrev4-2}
\bibliography{ref}

\newpage
\onecolumngrid
\appendix 
\clearpage
\renewcommand{\thesubsection}{\thesection\arabic{subsection}}
\renewcommand\thefigure{S\arabic{figure}}    
\setcounter{figure}{0} 
\renewcommand{\theequation}{\thesection\arabic{equation}}
\setcounter{equation}{0}
\makeatletter
\renewcommand{\p@subsection}{}
\makeatother

\section{Three-qutrit fixed marginals with distinct multipartite invariants}
\label{sec:sm_three-qutrit}

We give an explicit continuous family of pure states for which all
three one-party density matrices remain fixed, while the four-copy
invariant used in the main text changes. This example establishes
the algebraic distinction between bipartite spectral data and
multipartite local-unitary structure; it does not assume equilibration.
\vspace{0.3cm}
\paragraph{Normalized states and fixed complete marginals.}
Let $\epsilon_{012}=1$ and define the orthonormal states
\begin{equation}
\begin{aligned}
|\Phi\rangle&=\frac{1}{\sqrt3}\sum_{j=0}^{2}|jjj\rangle,~~~~~
|\Omega\rangle=\frac{1}{\sqrt6}
\sum_{i,j,k=0}^{2}\epsilon_{ijk}|ijk\rangle,\\
|\psi_\theta\rangle&=\cos\theta\,|\Phi\rangle
+\sin\theta\,|\Omega\rangle,
\qquad 0\leq\theta\leq\frac{\pi}{2}.
\end{aligned}
\label{eq:sm_qutrit_family}
\end{equation}
Their orthogonality follows because a nonzero Levi--Civita symbol
has three distinct indices, whereas the GHZ state has three equal
indices. In particular, $|\psi_\theta\rangle$ is normalized.

For party $A$, the endpoint marginals are
\begin{equation}
\begin{aligned}
\operatorname{Tr}_{BC}|\Phi\rangle\langle\Phi|
=\frac{\mathbb I_3}{3},\qquad~~~
\bigl(\operatorname{Tr}_{BC}|\Omega\rangle\langle\Omega|\bigr)_{ii'}
=\frac16\sum_{j,k}\epsilon_{ijk}\epsilon_{i'jk}
=\frac{\delta_{ii'}}3.
\end{aligned}
\label{eq:sm_qutrit_end_marginals}
\end{equation}
The interference term vanishes as a matrix, since
\begin{equation}
\bigl(\operatorname{Tr}_{BC}|\Phi\rangle\langle\Omega|\bigr)_{ii'}
=\frac1{\sqrt{18}}\sum_{j,k}
\delta_{ij}\delta_{ik}\epsilon_{i'jk}=0.
\label{eq:sm_qutrit_cross_trace}
\end{equation}
Its Hermitian conjugate also vanishes. The same calculation applies
to $B$ and $C$, so the complete one-party matrices obey
\begin{equation}
\rho_A(\theta)=\rho_B(\theta)=\rho_C(\theta)
=\frac{\mathbb I_3}{3}.
\label{eq:sm_qutrit_fixed_marginals}
\end{equation}
Consequently, $S_n(X)=\log3$ for every $n>0$, including the
von Neumann limit, and for $X=A,B,C$. Purity of $ABC$ also fixes
the nonzero spectra of the complementary two-party matrices.
Those two-party matrices themselves need not remain fixed.
\vspace{0.3cm}
\paragraph{Four-copy contraction by realignment.}
For a normalized pure state of $ABC$, set
\begin{equation}
R_{ii',jj'}=(\rho_{AB})_{ij,i'j'}.
\end{equation}
Tracing each $C$ index in the four-replica definition and then
collecting the $A$ and $B$ indices gives
\begin{equation}
\begin{aligned}
Z_2^{(3)}
&=\operatorname{Tr}\!\left[
\rho^{\otimes4}P_A((13)(24))P_B((12)(34))\right]=\operatorname{Tr}\!\left[(RR^\dagger)^2\right].
\end{aligned}
\label{eq:sm_qutrit_realignment}
\end{equation}
This expression is real and nonnegative and is invariant under
$U_A\otimes U_B\otimes U_C$.

For a direct calculation throughout the family, write
$x=\cos\theta/\sqrt3$ and $y=\sin\theta/\sqrt6$.
The amplitudes are
$\psi_{ijk}=x\delta_{ij}\delta_{ik}+y\epsilon_{ijk}$.
Using
$\sum_k\epsilon_{ijk}\epsilon_{i'j'k}
=\delta_{ii'}\delta_{jj'}-\delta_{ij'}\delta_{ji'}$
gives
\begin{equation}
\begin{aligned}
(\rho_{AB})_{ij,i'j'}={}&
x^2\delta_{ij}\delta_{i'j'}\delta_{ii'}
+y^2(\delta_{ii'}\delta_{jj'}-\delta_{ij'}\delta_{ji'})
+xy\left(\delta_{ij}\epsilon_{i'j'i}
+\delta_{i'j'}\epsilon_{iji'}\right).
\end{aligned}
\label{eq:sm_qutrit_rhoAB}
\end{equation}
Realigning these entries and taking the trace in
Eq.~\eqref{eq:sm_qutrit_realignment} yields the homogeneous polynomial
\begin{equation}
Z_2^{(3)}(\theta)
=3x^8+72x^4y^4+48x^2y^6+24y^8.
\label{eq:sm_qutrit_polynomial}
\end{equation}
Thus, putting $c_\theta=\cos\theta$ and $s_\theta=\sin\theta$,
\begin{equation}
Z_2^{(3)}(\theta)
=\frac{2c_\theta^8+12c_\theta^4s_\theta^4
+4c_\theta^2s_\theta^6+s_\theta^8}{54}.
\label{eq:sm_qutrit_Ztheta}
\end{equation}
Because the product of the three single-party purities is $1/27$,
the genuine multi-entropy in the convention of the main text is
\begin{equation}
\begin{aligned}
G(\theta)
&=-\frac12\log\!\left[27Z_2^{(3)}(\theta)\right]=\frac12\log\!\frac{2}{
2c_\theta^8+12c_\theta^4s_\theta^4
+4c_\theta^2s_\theta^6+s_\theta^8}.
\end{aligned}
\label{eq:sm_qutrit_Gtheta}
\end{equation}
\vspace{0.3cm}
\paragraph{Independent endpoint evaluations and interpretation.}
For $|\Phi\rangle$, the realigned matrix is
$R_\Phi=\frac13\sum_j|jj\rangle\langle jj|$.
It has three nonzero singular values, all equal to $1/3$, and hence
$Z_2^{(3)}(\Phi)=3(1/3)^4=1/27$.
For $|\Omega\rangle$,
\begin{equation}
\rho_{AB}^{\Omega}=\frac{\mathbb I-\mathbb F}{6},
\qquad
R_\Omega=\frac{|I\rangle\!\rangle\langle\!\langle I|-\mathbb F}{6},
\end{equation}
where $\mathbb F|ij\rangle=|ji\rangle$ and
$|I\rangle\!\rangle=\sum_j|jj\rangle$.
Since $\langle\!\langle I|I\rangle\!\rangle=3$,
$R_\Omega$ has one singular value $1/3$ and eight singular
values $1/6$. Therefore,
\begin{equation}
\begin{aligned}
Z_2^{(3)}(\Omega)&=\left(\frac{1}{3}\right)^4+8\left(\frac{1}{6}\right)^4=1/54,\qquad
G(0)=0,\qquad G\left(\frac{\pi}{2}\right)=\frac12\log2.
\end{aligned}
\label{eq:sm_qutrit_endpoints}
\end{equation}
These values agree with Eqs.~\eqref{eq:sm_qutrit_Ztheta}
and~\eqref{eq:sm_qutrit_Gtheta}; for example,
$G(\pi/4)=\frac12\log(32/19)$.
Since an LU invariant differs between these states, fixing all
three one-party matrices cannot fix the multipartite LU class.
One nonconstant invariant suffices to establish this distinction;
it need not be a complete classifier or an entanglement monotone.

\section
{Type I -- From Entanglement Plateaus to Spectrum-Invisible Dynamics}
\label{sec:sm_typeI}

We collect the definitions and calculations behind {the three} {\bf Type I} examples
in the main text. The tilted-field chain establishes the comparison protocol
for different information measures; the kicked-Ising chain supplies an exact
fixed-marginal interval; and the Haar Choi circuit permits an independent
control of the complete marginal spectra. Throughout, the three parties
form one globally pure state.

\subsection{Tilted-field Ising: spectral and higher-replica diagnostics}\label{sec:sm_tilted}

\paragraph{Hamiltonian, states, and partitions.}
We use the open-chain Hamiltonian
\begin{equation}
 H_{\mathrm{TI}}(J)=J\sum_{j=1}^{N-1}Z_jZ_{j+1}
 +1.05\sum_{j=1}^N X_j+0.5\sum_{j=1}^N Z_j+0.37Z_1,
 \label{eq:sm_tilt_H}
\end{equation}
{with $J=1$, $\hbar=1$, and three equal contiguous regions $A|B|C$ for $N=9,12,15,18$. For each $N$ we independently
prepare five equatorial product states,}
\begin{equation}
 |\psi_0\rangle=\bigotimes_{j=1}^N
 \frac{|0\rangle+e^{i\phi_j}|1\rangle}{\sqrt2},
 \qquad \sum_{j=1}^N\cos\phi_j=0.
 \label{eq:sm_tilt_initial}
\end{equation}
{In numerical calculations, to sample these states, we draw $N-1$ independent uniform phases, reject a draw unless $|\sum_{j<N}\cos\phi_j|\leq1$, choose the last phase to enforce the constraint (with a randomly selected inverse-cosine branch), and randomly permute the sites.
Since each site has $\langle Z_j\rangle=0$ and the initial state is a product, all sampled states obey $\langle H_{\mathrm{TI}}\rangle =1.05\sum_j\cos\phi_j=0$, although their energy variances can differ. We evolve $|\psi(t)\rangle=e^{-iH_{\mathrm{TI}}t}|\psi_0\rangle$.}

\vspace{0.3cm}
\paragraph{Operational equilibration times.}

{Each trajectory is sampled at $t_i=0,0.25,\ldots,60$ (241 points), and reference values denoted by an overbar are time averages over $48\leq t_i\leq60$ (49 points), computed \emph{separately for each initial state}. For ordinary R\'enyi entropies we use
\begin{equation}
 D_{S_n}(t)=\bigg[\frac13\sum_{X=A,B,C}
 \bigl(S_n(X,t)-\overline{S_n(X)}\bigr)^2\bigg]^{1/2}.
 \label{eq:sm_tilt_entropy_deviation}
\end{equation}
The genuine tripartite four-replica invariant is
$G=GM_2^{(3)}$, with
$D_{GM_2}(t)=|G(t)-\overline G|$.
For the complete single-region Schmidt spectra, write $p_i^X(t)$ for the eigenvalues of $\rho_X(t)$ sorted in descending order and $\overline p_i^X$ for their individually time-averaged values in the reference window. We define
\begin{equation}
 D_{\mathrm{spec}}(t)=\left\{\frac13\sum_{X=A,B,C}
 \left[\frac12\sum_i|p_i^X(t)-\overline p_i^X|\right]^2
 \right\}^{1/2}.
 \label{eq:sm_spec_dist}
\end{equation}
The spectral distance is insensitive to Schmidt-basis rotations. We also study the min-entropy $S_\infty(X,t)=-\ln p_{\max}^X(t)$, which probes only the largest Schmidt eigenvalue and may therefore exhibit a different operational equilibration time from other R\'enyi entropies. Its deviation is defined by Eq.~\eqref{eq:sm_tilt_entropy_deviation} with $n=\infty$.

For each entanglement quantity, we apply five-point centered smoothing and define $t_{90}$ as the first time the smoothed deviation remains below $10\%$ of its unsmoothed initial value for nine consecutive samples (two time units). Crossing times are determined separately for each initial state before averaging. They characterize operational finite-window plateaus rather than asymptotic relaxation times.
}

\begin{figure}[h!]
\centering
\includegraphics[width=0.43\columnwidth]
{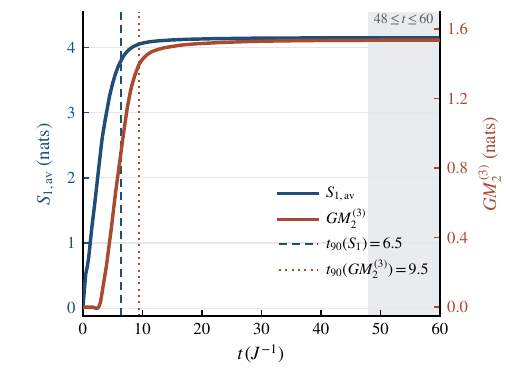}
\caption{\small 
Representative evolution of the tilted-field Ising chain
at $N=18$ for a single initial state.
The mean single-party von Neumann entropy
$S_{1,\mathrm{av}}=\frac13\sum_{X=A,B,C}S_1(X)$
and $GM_2^{(3)}$ are shown on the left and right
vertical axes, respectively.
The vertical lines indicate the operational crossing
times $t_{90}(S_1)=6.5$ and
$t_{90}(GM_2^{(3)})=9.5$.
The shaded region marks the late-time reference
window $48\leq t\leq60$.
}
\label{fig:sm_n18_ee_gm}
\end{figure}

\vspace{0.3cm}
\paragraph{Ordinary entropies, full spectra, and multi-entropy.} 
The mean $10\%$ crossing times from five initial states at each size are
\begin{center}
\begingroup
\setlength{\tabcolsep}{10pt}
\renewcommand{\arraystretch}{1.3}
\begin{tabular}{c|rrrrrr}
\hline \hline
$N$ & $t_{90}(S_1)$ & $t_{90}(S_2)$
    & $t_{90}(S_3)$ & $t_{90}(GM_2^{(3)})$
    & $t_{90}(D_{\rm spec})$ & $t_{90}(S_\infty)$\\ \hline
$9$  & 3.45 & 3.95 & 4.25 & 5.35 & 4.45 & 5.00\\
$12$ & 4.75 & 5.65 & 6.00 & 7.15 & 6.75 & 7.40\\
$15$ & 5.45 & 6.25 & 6.70 & 7.75 & 7.85 & 8.25\\
$18$ & 6.35 & 7.30 & 7.85 & 9.05 & 9.85 & 10.70\\ 
\hline \hline
\end{tabular}
\endgroup
\end{center}
All reported $t_{90}$ values are means over five independently prepared initial states, rather than crossing times common to all five realizations. At $N=18$, the sample standard deviations of $t_{90}(S_1)$, $t_{90}(GM_2^{(3)})$, $t_{90}(D_{\mathrm{spec}})$, and $t_{90}(S_\infty)$ are $0.22$, $0.37$, $0.99$, and $1.54$, respectively. A representative trajectory, together with its operational crossing times and late-time reference window, is shown in Fig.~\ref{fig:sm_n18_ee_gm}. As a threshold check, the corresponding mean crossing times at $5\%$ are $7.60$, $10.15$, $12.65$, and $14.20$. The dependence of these crossing times on both diagnostic and threshold is why we regard them as operational plateau times rather than microscopic decay rates.

\vspace{0.3cm}
\paragraph{Finite-window effective rates.}
As a separate, descriptive check, for each initial state we fit $\ln D(t)=a-\gamma t$ to the \emph{raw}, positive distances $D_{\mathrm{EE}}=D_{S_1}$ and $D_{\mathrm{GM}}=D_{GM_2}$ over $t=2,2.25,\ldots,20$. At $N=18$, averaging the five independently fitted slopes gives
\begin{equation}
\begin{split}
 \overline\gamma_{\mathrm{EE}}=0.4014\pm0.0864,~~~~~~~~~ \overline\gamma_{\mathrm{GM}}=0.4157\pm0.0915,
\end{split}
\label{eq:sm_tilt_effective_rates}
\end{equation}
where the uncertainties are sample standard deviations; the mean fit coefficients of determination are $R^2_{\mathrm{EE}}=0.961$ and $R^2_{\mathrm{GM}}=0.966$. These fits include the early-time transient and use the $48$--$60$ references, not a late-time asymptotic regime. The comparable fitted slopes indicate that the deviations of EE and GM decay at similar effective rates over this finite window, despite their different operational equilibration times. However, this does not establish that they are governed by the same microscopic relaxation mode or determine an asymptotic spectral gap.

\vspace{0.3cm}
\paragraph{Higher-replica multipartite invariants.}
To test more than the four-replica representative $G=GM_2^{(3)}$,
we calculate the standard $n^2$-replica tripartite multi-entropies.
They are not obtained by simply raising the four-replica matrix
contraction to a higher power. Label the $n^2$ copies by
$(r,s)\in\mathbb Z_n^2$ and introduce two independent shifts,
\begin{equation}
 \pi_A(r,s)=(r+1,s),\qquad
 \pi_B(r,s)=(r,s+1),\qquad
 \pi_C=\mathrm{id},
 \label{eq:sm_replica_grid}
\end{equation}
with indices understood modulo $n$. For a normalized pure state,
\begin{equation}
\begin{split}
 Z_n^{(3)}=&\sum_{\{a_{rs},b_{rs},c_{rs}\}}
 \prod_{r,s=0}^{n-1}
 \psi_{a_{rs}b_{rs}c_{rs}}
 \psi^*_{a_{r+1,s}\,b_{r,s+1}\,c_{rs}},\\
 S_n^{(3)}=&\frac{\log Z_n^{(3)}}{n(1-n)},\qquad
 GM_n^{(3)}=S_n^{(3)}-\frac12\sum_{X=A,B,C}S_n(X).
\end{split}
\label{eq:sm_replica_gmn}
\end{equation}
The $n=2$ contraction is checked against the independent realignment
formula used in Sec.~II, and the higher-$n$ implementation is tested
against analytic product and GHZ-state values. Standard $GM_3^{(3)}$
requires \emph{nine} replicas, while $GM_4^{(3)}$ requires
\emph{sixteen}; simply replacing
$\operatorname{Tr}(RR^\dagger)^2$ by a higher matrix power does not
compute these quantities.

At $N=9$, five complete $GM_3^{(3)}$ trajectories (241 time samples each) give
\begin{equation}
t_{90}(GM_2^{(3)}) = 5.35\pm0.52,\qquad
t_{90}(GM_3^{(3)}) = 10.20\pm4.64,
\label{eq:sm_gm3_times}
\end{equation}
where uncertainties are state-to-state sample standard deviations. For the five initial states, the individual $GM_3^{(3)}$ crossing times are $9.50$, $18.00$, $6.00$, $10.00$, and $7.50$, respectively, each exceeding the corresponding $GM_2^{(3)}$ crossing time. The mean $GM_3^{(3)}$ crossing time also exceeds the $S_\infty$ mean of $5.00\pm0.64$. However, the estimated $95\%$ confidence interval for the mean paired difference $t_{90}(GM_3^{(3)})-t_{90}(S_\infty)$ is $[-0.53,10.93]$, so these data do not establish a universal ordering. Moreover, all five $GM_3^{(3)}$ trajectories leave their $10\%$ deviation bands at least once after the first sustained crossing. These crossing times therefore characterize finite-window operational plateaus rather than irreversible relaxation.

For higher replica indices, selected exact calculations at $N=6$ yield
\begin{equation}
\begin{split}
GM_4^{(3)}(10)=0.11479,~~~ 
GM_4^{(3)}(20)&=0.08255,~~~
GM_4^{(3)}(36)=0.07434,\\
GM_5^{(3)}(10)&=0.08847.
\end{split}
\label{eq:sm_gm4_sparse}
\end{equation}
These sparse $4^2$- and $5^2$-replica calculations provide additional checks of the higher-replica contractions but neither determine reliable plateau-crossing times nor establish a hierarchy of equilibration times with replica indices.


\subsection{Self-dual kicked Ising: exact marginal plateau and replica readout}
\label{sec:sm_sdki}

\paragraph{Why the complete marginals become stationary.}
We use the periodic Floquet operator in the main text and
$|\psi(0)\rangle=|0\rangle^{\otimes N}$, with $N=3\ell$.
The diagonal first segment changes this initial state only by a global
phase, so after one complete period
\begin{equation}
|\psi(1)\rangle\simeq
\bigotimes_{j=1}^{N}\frac{|0\rangle_j-i|1\rangle_j}{\sqrt2}.
\label{eq:sm_sdki_first_period}
\end{equation}
This is a transverse separating state. The standard dual-unitary
separating-state contraction gives, on the infinite chain,
\begin{equation}
\operatorname{Tr}\rho_X(\tau)^2
=2^{-\min(2\tau,\ell)}
\end{equation}
for a block of $\ell$ sites
\cite{BertiniKosProsen2019Entanglement}.
For the polarized initial state used here, its effective time is
$\tau=t-1$, since the first period creates no entanglement.
Each further period extends a backward cone by at most one site on
either side. The same folded contraction applies on the finite ring
while the two outer fronts have not passed through the complement,
namely $2\tau\leq N-\ell$; the endpoint is included with the stated
discrete gate ordering. Combining this condition with
$2\tau\geq\ell$ gives
\begin{equation}
\operatorname{Tr}\rho_X(t)^2=2^{-\ell},\qquad
1+\left\lceil\frac\ell2\right\rceil\leq t\leq\ell+1,
\quad X=A,B,C.
\label{eq:sm_sdki_purity}
\end{equation}
For any normalized density matrix of dimension $d_X=2^\ell$,
\begin{equation}
\left\|\rho_X-\frac{\mathbb I}{d_X}\right\|_2^2
=\operatorname{Tr}\rho_X^2-\frac1{d_X}.
\label{eq:sm_sdki_saturation}
\end{equation}
Thus the purity identity proves the complete matrix equality
$\rho_X=2^{-\ell}\mathbb I$ throughout this finite window.
The contraction does not assert permanent stationarity after the
causal fronts wrap around the ring.

\vspace{0.3cm}
\paragraph{Independent state-vector checks.}
The preceding identities can also be tested without employing a
tensor-network simplification. First apply the diagonal factor
\begin{equation}
 \psi_{\boldsymbol z}\longmapsto
 \exp\!\left[-i\left(\frac\pi4\sum_j z_jz_{j+1}
 +h\sum_jz_j\right)\right]\psi_{\boldsymbol z},
 \qquad z_j=\pm1.
 \label{eq:sm_sdki_diag}
\end{equation}
Then apply the on-site rotations
$e^{-i\pi X_j/4}$, site by site, and form each
$\rho_X=\operatorname{Tr}_{\bar X}|\psi(t)\rangle\langle\psi(t)|$.
At $h=0.15$ the following windows were independently checked:
\begin{center}
\setlength{\tabcolsep}{10pt}
\renewcommand{\arraystretch}{1.3}
\begin{tabular}{c|c|c|c}
\hline
$N$ & $\ell$ & exact-marginal times
& largest $\max_X\|\rho_X-2^{-\ell}\mathbb I\|_2$\\\hline
6 &2 &$t=2,3$ & $<2\times10^{-15}$\\
12&4 &$t=3,4,5$ & $<3\times10^{-15}$\\
18&6 &$t=4,5,6,7$ & $<3\times10^{-15}$\\\hline
\end{tabular}
\end{center}
The quoted residuals reflect floating-point roundoff. At the first integer times beyond these intervals, $t=4,6,8$ for $N=6,12,18$, respectively, the maximum Frobenius deviations increase to $0.05980$, $0.31933$, and $0.14124$. These results confirm that the exact marginal plateaus are restricted to finite time windows, with nonzero deviations appearing thereafter.

\vspace{0.3cm}
\paragraph{Multipartite contraction.}
For a pure $ABC$ state let
$R_{aa',bb'}=(\rho_{AB})_{ab,a'b'}$ denote the
realignment of $\rho_{AB}$. Then
\begin{equation}
 Z_2^{(3)}=\operatorname{Tr}[(RR^\dagger)^2],\qquad
 G=-\frac12\log Z_2^{(3)}
 -\frac12\sum_{X=A,B,C}S_2(X).
 \label{eq:sm_sdki_realign}
\end{equation}
Related realignment descriptions of replica invariants are discussed in Ref.~\cite{BerthiereGaudin2026}.
The $N=12$, $\ell=4$, $h=0.15$ evolution gives

\begin{center}
\setlength{\tabcolsep}{10pt}
\renewcommand{\arraystretch}{1.3}
\begin{tabular}{c|c|c}
\hline \hline
$t$ & $\max_X\|\rho_X-\mathbb I_{16}/16\|_2$
 & $G(t)$ \\\hline
3 & $2.77\times10^{-16}$ & $0$\\
4 & $2.29\times10^{-16}$ & $0.238767159$\\
5 & $2.37\times10^{-16}$ & $0.113581124$\\
\hline \hline
\end{tabular}
\end{center}

The norm and $G$ are computed independently from the same full
state vector. Since $G$ is unchanged under any
$U_A\otimes U_B\otimes U_C$, its variation within an interval of
identical one-party matrices establishes motion between different
multipartite LU structures. The nonmonotonicity reflects coherent
finite-system dynamics; this example supplies no separate slow rate.


\subsection{Haar Choi circuits: from replica plateaus to full-marginal decoupling}
\label{sec:sm_haar_typeI}

\paragraph{Circuit, partition, and annealed moments.}

\begin{figure}[t]
  \centering
  \includegraphics[width=0.83\columnwidth]{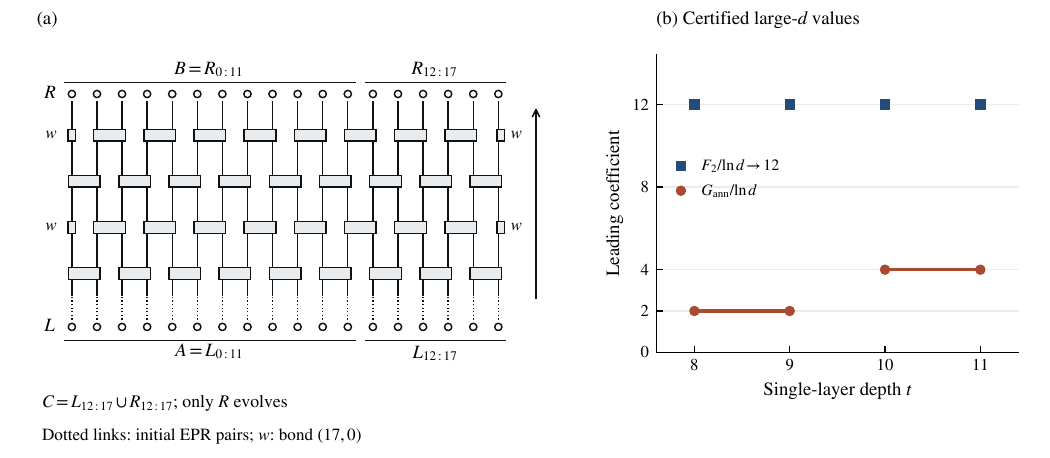}
  \caption{\small Certified annealed separation in a Haar Choi circuit.
  (a) Aligned intervals $A\subset L$ and $B\subset R$, each of size $\ell$,
  and their common complement $C$.
  (b) At $N=18$, $\ell=12$, $m=6$, the certified depths $t=8,10$
  obey $F_4=(40,44)\log d+O(d^{-2})$ and $F_2=12\log d+o(1)$;
  hence $G_{\rm ann}=(2,4)\log d+O(d^{-2})$.
  The marginal deviations are $\Delta_C(8)=4836d^{-4}+O(d^{-6})$
  and $\Delta_C(10)=61479d^{-8}+O(d^{-10})$.
  Odd layers leave the invariants unchanged; no unverified smooth
  growth curve or typical-sample assertion is inferred.}
  \label{fig:typeI_haar}
\end{figure}


For even $N$, alternate independent Haar $U(d^2)$ gates on the
nearest-neighbor bonds $(0,1),(2,3),\ldots$ and
$(1,2),(3,4),\ldots,(N-1,0)$ of a periodic ring.
Let $I=\{\ell,\ldots,N-1\}$ contain $m=N-\ell$ sites,
and write the normalized Choi state as
\begin{equation}
 |U(t)\rangle=d^{-N/2}\sum_{\boldsymbol i,\boldsymbol j}
 U_{\boldsymbol j\boldsymbol i}(t)
 |\boldsymbol i\rangle_L|\boldsymbol j\rangle_R.
 \label{eq:sm_haar_choi}
\end{equation}
We partition the complete pure state into
$A=L_{I^c}$, $B=R_{I^c}$ and $C=L_I\cup R_I$.
The dimensions are $D_A=D_B=d^\ell$ and $D_C=d^{2m}$.
Figure~\ref{fig:typeI_haar} summarizes the circuit partition
and the certified large-$d$ separation at $t=8,10$.
For \emph{every} unitary, before ensemble averaging,
\begin{equation}
 \rho_A=\frac{\mathbb I_A}{d^\ell},\qquad
 \rho_B=\frac{\mathbb I_B}{d^\ell}.
 \label{eq:sm_haar_AB}
\end{equation}
The only nontrivial ordinary spectrum is that of $\rho_C$.
Define
\begin{equation}
 \begin{split}
 P_2(t)&=\mathbb E_U\operatorname{Tr}\rho_C(t)^2,
 \qquad F_2(t)=-\log P_2(t),
 \qquad 
 Z_4(t)=\mathbb E_U Z_2^{(3)}(t),
 \qquad F_4(t)=-\log Z_4(t),\\
 G_{\mathrm{ann}}(t)&=\frac12[F_4(t)-F_2(t)]
 -\ell\log d.
 \end{split}
 \label{eq:sm_haar_annealed}
\end{equation}
This $G_{\mathrm{ann}}$ is constructed from logarithms of averaged
replica moments and is \emph{not} generally
$\mathbb E_U[GM_2^{(3)}]$.

\vspace{0.3cm}
\paragraph{Finite-layer replica costs and signed four-copy result.}
The Haar average is evaluated in $S_r$ permutation variables via
Weingarten calculus. Geometric domain-wall minima are candidates for
large-$d$ costs, but the four-copy local coefficients carry signs.
The former formulas $2\min(t,\ell,m)$ and
$2\min(\ell+t,2\ell,N+m)$ cannot be used as fixed-geometry,
layer-resolved expansions with $O(1)$ error.

At the chosen aligned even partition, each odd-numbered brickwork
layer consists only of intra-party unitaries, giving the exact identities
\begin{equation}
F_r(2p+1)=F_r(2p),\qquad r=2,4.
\label{eq:sm_haar_odd_layers}
\end{equation}
In particular $F_2(0)=F_2(1)=0$ and
$F_4(0)=F_4(1)=2\ell\log d=24\log d$ at $\ell=12$.
The previous continuous first-layer costs would differ by
$2\log d$, not by an $O(1)$ remainder.

For the fixed $N=18$, $\ell=12$, $m=6$
circuit, the exact signed-network calculation,
combined with finite-graph integer certification, yields
\begin{equation}
\begin{aligned}
Z_4(8)&=d^{-40}\,[1+O(d^{-2})],~~~~~~~
Z_4(10)=d^{-44}\,[1+O(d^{-2})].
\end{aligned}
\label{eq:sm_haar_certmoments}
\end{equation}
Neither this finite-geometry result nor odd-layer constancy
is a statement about typical-sample GM, a general depth formula,
or the thermodynamic limit at fixed $d$.

\vspace{0.3cm}
\paragraph{A stronger full-marginal criterion.}
Let
\begin{equation}
 \Delta_C(t)=D_C\,\mathbb E_U\operatorname{Tr}\rho_C(t)^2-1
 =D_C\,\mathbb E_U\left\|
 \rho_C(t)-\frac{\mathbb I_C}{D_C}\right\|_2^2.
 \label{eq:sm_haar_delta}
\end{equation}
The matrix norm bound $\|M\|_1\leq\sqrt{D_C}\,\|M\|_2$,
followed by Cauchy--Schwarz for the ensemble average, gives
\begin{equation}
 \mathbb E_U\left\|\rho_C(t)-\frac{\mathbb I_C}{D_C}\right\|_1
 \leq\sqrt{\Delta_C(t)},
 \label{eq:sm_haar_bound}
\end{equation}
and Markov's inequality gives, for $\epsilon>0$,
\begin{equation}
 \mathbb P_U\!\left(
 \left\|\rho_C(t)-\frac{\mathbb I_C}{D_C}\right\|_1>
 \epsilon\right)\leq\frac{\Delta_C(t)}{\epsilon^2}.
 \label{eq:sm_haar_prob}
\end{equation}
Consequently $\Delta_C(t)\to0$ is sufficient for the complete
marginal to converge to the maximally mixed state in average trace
norm and in probability, not merely in leading entropy density.
\vspace{0.3cm}
\paragraph{Exact two-replica operator-memory identity.}
Choose normalized generalized Pauli (Weyl) bases
$\{P_\alpha\}$ and $\{W_\beta\}$ on the $m$ sites of $I$,
with $\operatorname{Tr}P_\alpha^\dagger P_{\alpha'}=
\operatorname{Tr}W_\beta^\dagger W_{\beta'}=d^m\delta_{\alpha\alpha'}$.
The Choi identity and operator-basis Parseval relation yield
\begin{equation}
 \begin{split}
 \Delta_C(t)=\sum_{\alpha\ne0,\,\beta\ne0}
 \mathbb E_U\Big|d^{-N}\operatorname{Tr}\big[
 W_\beta^\dagger U(t)
 (P_\alpha^{\mathsf T}\otimes\mathbb I_{I^c})
 U(t)^\dagger\big]\Big|^2.
 \end{split}
 \label{eq:sm_haar_memory}
\end{equation}
Here $W_\beta$ is implicitly embedded into the full output ring.
Thus $\Delta_C$ is the total squared memory of input operators
initially supported in $I$ that remain supported in the same output
interval after random-circuit spreading.
\vspace{0.3cm}
\paragraph{Exact Haar support-transfer rule.}
Write $Q=d^2$ and encode the support of each Pauli string by a
bitmask $S\subseteq\{0,\ldots,N-1\}$. For a two-site Haar gate, an
incoming $00$ support stays $00$. Every nonempty incoming support
is redistributed according to~\cite{NahumVijayHaah2018}
\begin{equation}
 p_{10}=p_{01}=\frac1{Q+1},\qquad
 p_{11}=\frac{Q-1}{Q+1},
 \label{eq:sm_haar_support_rule}
\end{equation}
which follows from the multiplicities $Q-1,Q-1,(Q-1)^2$
of traceless two-site generalized Pauli strings.
Set
\begin{equation}
 w_0(S)=\begin{cases}
 (Q-1)^{|S|},&\varnothing\ne S\subseteq I,\\[0.3cm]
 0,&\text{otherwise},
 \end{cases}
 \quad w_t=\mathcal T_t\cdots\mathcal T_1w_0,
 \label{eq:sm_haar_initial}
\end{equation}
where $\mathcal T_j$ is the product of local support-transfer
maps in layer $j$. The exact finite-$d$ recurrence is
\begin{equation}
 \Delta_C(t)=\sum_{\varnothing\ne S\subseteq I}w_t(S).
 \label{eq:sm_haar_recursion}
\end{equation}
This finite-dimensional recurrence involves only nonnegative
weights and is exact at any finite $d$; it provides an independent
check of full-marginal equilibration, separate from the leading
four-copy large-$d$ calculation.
At $t=0$, $\Delta_C(0)=Q^m-1$, as required by the Choi state of
the identity channel.
\vspace{0.3cm}
\paragraph{Boundary layer and explicit decoupling.}
The transfer process detects a short stroboscopic boundary layer:
ordinary leading-cost saturation does not automatically imply
$\Delta_C\to0$ in the same brickwork layer. For example, the exact
$N=6,m=2$ result
\begin{equation}
 \Delta_C(2)=\Delta_C(3)
 =\left(\frac{2Q}{Q+1}\right)^2-1
 \longrightarrow3\quad(d\to\infty)
 \label{eq:sm_haar_boundary_layer}
\end{equation}
shows that complete-spectrum decoupling fails at the first
leading-cost plateau, despite its apparently saturated $F_2$.

For the stronger example used in the main text, $N=18,\ell=12,m=6$,
exact evaluation of the positive transfer recurrence gives
\begin{equation}
 \begin{split}
 \Delta_C(7)&=399+O(d^{-2}),\\
 \Delta_C(8)&=4836\,d^{-4}+O(d^{-6}),\\
 \Delta_C(10)&=61479\,d^{-8}+O(d^{-10}).
 \end{split}
 \label{eq:sm_haar_large_d_coefficients}
\end{equation}
The coefficients follow by expanding the positive recurrence in
$1/d$; they are not extrapolated from a fit. In particular,
$\mathbb E\|\rho_C(8)-\mathbb I/d^{12}\|_1=O(d^{-2})$
and
$\mathbb E\|\rho_C(10)-\mathbb I/d^{12}\|_1=O(d^{-4})$.
The $A$ and $B$ marginals are already fixed exactly.
At these same two depths, the exact signed four-copy result gives
\begin{equation}
 G_{\mathrm{ann}}(10)-G_{\mathrm{ann}}(8)
 =2\log d+O(d^{-2}).
 \label{eq:sm_haar_final}
\end{equation}
The limit here is $d\to\infty$ at fixed $N,\ell,m,t$.
The controlled full-marginal statement concerns average trace norm
(and, by the bound above, convergence in probability). The
four-copy observable is the annealed quantity $G_{\mathrm{ann}}$,
not the average of samplewise logarithms.

\begin{figure}[h!]
\centering
\includegraphics[width=0.82\linewidth]
{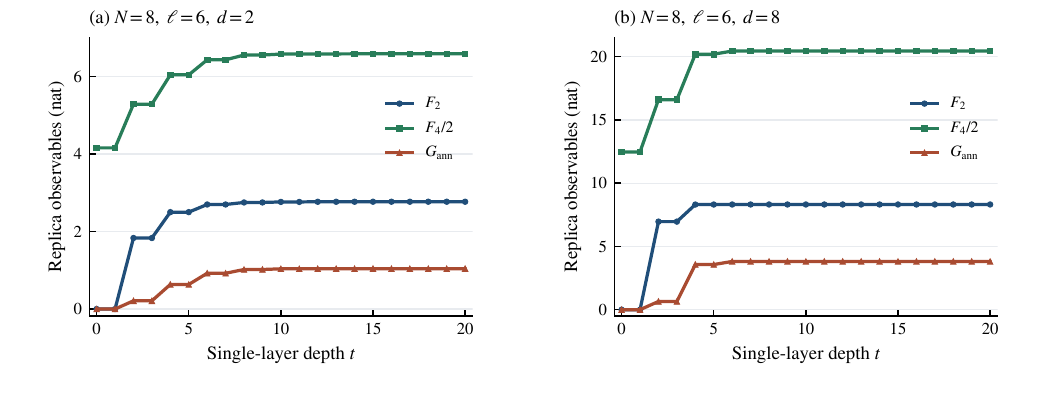}
\caption{ \small
Annealed replica observables in the Haar Choi circuit with
$N=8$, $\ell=6$, and $m=2$, for $d=2$ and $d=8$.
Blue, green, and brown curves show $F_2$, $F_4/2$,
and $G_{\mathrm{ann}}$, respectively, with
$G_{\mathrm{ann}}=(F_4-F_2)/2-\ell\log d$.
Time is measured in individual brickwork layers.
}
\label{fig:sm_haar_finite_archived}
\end{figure}

For comparison, Fig.~\ref{fig:sm_haar_finite_archived} shows the
finite-$d$ annealed observables for $N=8$, $\ell=6$, $m=2$
at $d=2,8$. This smaller-geometry illustration is separate
from the certified fixed-geometry large-$d$ result above.


\subsection{Leading asymptotics of the signed four-replica contraction}
\label{sec:sm_haar_noncancellation}

\paragraph{Exact local Haar weights.}
For $g\in S_4$, define
$|g|=4-\#\operatorname{Cycles}(g)$.
The normalized one-site and two-site Gram matrices are
\begin{equation}
\begin{aligned}
C_{gh}&=d^{-|g^{-1}h|},\qquad 
H_{gh}=d^{-2|g^{-1}h|}.
\end{aligned}
\end{equation}
The exact local transfer coefficient, obtained from Haar
Weingarten integration~\cite{CollinsSniady2006}, is
\begin{equation}
J(c;a,b)=\sum_{\tau\in S_4}
[H^{-1}]_{c\tau}C_{\tau a}C_{\tau b}.
\label{eq:sm_haar_J}
\end{equation}
Writing $q=d^2$, the inverse satisfies
$[H^{-1}]_{c\tau}
=q^4\operatorname{Wg}_q(c^{-1}\tau)$.
The Weingarten functions for cycle types
$1^4$, $2\,1^2$, $2^2$, $3\,1$, and $4$ have the common
denominator $q^2(q^2-1)(q^2-4)(q^2-9)$ and respective numerators
\begin{equation}
\begin{aligned}
&q^4-8q^2+6,\qquad -q(q^2-4),\qquad
q^2+6,\qquad 2q^2-3,\qquad -5q.
\end{aligned}
\end{equation}
Thus the local weights are exact rational functions.
In particular, $C_{\tau a}^2=H_{\tau a}$ implies
$J(c;a,a)=\delta_{ca}$.
The weights can have either sign; for example,
\begin{equation}
\begin{aligned}
J((234);e,(243))
&=-\frac{2(d^2-1)}
{(d^2-2)(d^2+1)(d^2+2)}.
\end{aligned}
\end{equation}
For every nonzero local weight, write
\begin{equation}
J_f=h_f d^{-v_f}\bigl[1+O(d^{-2})\bigr],
\end{equation}
where $v_f$ is its leading exponent and $h_f$ its signed
integer leading coefficient. Identically vanishing rational
functions are excluded from the sum.
\vspace{0.3cm}
\paragraph{Finite circuit and exponent bound.}
For $N=18$ and $\ell=12$, each brickwork layer has nine gates.
Set $\alpha=(13)(24)$ and $\beta=(12)(34)$.
The first layer fixes six gate labels to $\beta$ and the
remaining three to $e$. For layer $t\geq2$, gate $j$ receives
the preceding gate labels $(j,j+1)$ when $t$ is even and
$(j-1,j)$ when $t$ is odd, with gate indices modulo nine.
At the final physical boundary, set $a_i=\alpha$ for
$0\leq i<12$ and $a_i=e$ otherwise. For even depth $T$,
\begin{equation}
B_T=\sum_{j=0}^{8}\left(
|\sigma_{T,j}^{-1}a_{2j+1}|+
|\sigma_{T,j}^{-1}a_{2j+2}|\right),
\end{equation}
where physical indices are read modulo 18.
The normalized four-replica contraction is
\begin{equation}
Z_4(T)=\sum_{\{\sigma\}}d^{-B_T}\prod_fJ_f.
\end{equation}
For each contributing configuration, define
\begin{equation}
E=B_T+\sum_fv_f,\qquad
n_-=\sum_f\mathbf{1}_{\{h_f<0\}}.
\end{equation}
Here $n_-$ counts the local negative leading coefficients.

Let $b_i(s)$ be the contribution of node $i$ to $B_T$ when
its permutation label is $s$, and set $b_i=0$ away from the
final boundary. Integer node values $z_i$ and incidence
functions $p_{fi}(s)$ define the residuals
\begin{equation}
\begin{aligned}
u_i(s)
=b_i(s)-z_i+\sum_{f\ni i}p_{fi}(s),\qquad
r_f(\boldsymbol{s}_f)
=v_f(\boldsymbol{s}_f)
-\mathbf{1}_{\{h_f(\boldsymbol{s}_f)<0\}}-\sum_{i\in f}p_{fi}(s_i).
\end{aligned}
\label{eq:sm_haar_integer_dual}
\end{equation}
For $T=8$ and $T=10$, the finite-graph calculation yields
integer potentials for which all node and factor residuals
are nonnegative, with
\begin{equation}
\sum_i z_i=\kappa_T,\qquad
\kappa_8=40,\qquad \kappa_{10}=44.
\end{equation}
Summing the residuals cancels every incidence function and yields
\begin{equation}
E-n_--\kappa_T
=\sum_i u_i(s_i)+\sum_f r_f(\boldsymbol{s}_f)\geq0.
\label{eq:sm_haar_integer_bound}
\end{equation}

\vspace{0.3cm}
\paragraph{Absence of leading-order cancellation.}
Any configuration with $E=\kappa_T$ must have $n_-=0$ and
vanishing node and factor residuals. The zero-residual
constraints admit exactly one configuration at each of
$T=8$ and $T=10$. Direct substitution gives
$(\sum_fv_f,B_T)=(36,4)$ and $(40,4)$, respectively.
Every local leading coefficient in these configurations is
$+1$, so the leading coefficient of each complete contraction
is also $+1$.

The exact local exponents satisfy
$v(c;a,b)\equiv|a^{-1}b|\pmod 2$.
Every internal gate label occurs twice as a parent, and
$B_T$ is even for the chosen boundary labels. Hence all
total exponents are even. Together with the local expansions,
this gives
\begin{equation}
\begin{aligned}
Z_4(8)&=d^{-40}\bigl[1+O(d^{-2})\bigr],\qquad 
Z_4(10)=d^{-44}\bigl[1+O(d^{-2})\bigr].
\end{aligned}
\end{equation}
This establishes Eq.~\eqref{eq:sm_haar_certmoments}.
Combined with the two-replica decoupling result, it yields
Eq.~\eqref{eq:sm_haar_final}. These statements concern
$d\to\infty$ at fixed circuit size and depth, and the
multipartite quantity is $G_{\mathrm{ann}}$.


\section{Type II: Hydrodynamic Modes and Distributed Logical Information}
\label{sec:sm_typeII}


We collect the calculation details for the {\bf type II} examples in the main text. We first derive the charge-diffusion timescale of the $U(1)$-conserving circuit as an independent reference for the entanglement-growth diagnostics. We then analyze the five-qubit static code, deriving its third-order effective logical Hamiltonian and the separation between fast syndrome dressing and slow coherent multipartite dynamics. Next, we study the same code coupled to a locally scrambling nine-qubit bath, establishing the exact code-share decoupling identity, deriving the averaged logical channel, and detailing the paired-circuit numerical results and auxiliary four-copy Haar-isometry approximation. Finally, we present a complementary, exactly solvable code-preserving Bell-pair collision circuit, including its exact four-copy result and tunable $\lambda^{-2}$ relaxation timescale.

\subsection{$U(1)$-preserving circuit: exact diffusion and finite-size information diagnostics}
\label{sec:sm_typeII_hydro}

\paragraph{Charge-conserving local gates.}
Consider $N$ spin-$\frac12$ sites with open boundaries and occupation
$n_j=(1-Z_j)/2$. Every two-site random gate conserves
$n_j+n_{j+1}$, acting in the sectors of charges zero $|00\rangle$, one $\{|01\rangle,|10\rangle\}$, and two $|11\rangle$ as
\begin{equation}
 U_{j,j+1}=e^{i\phi_0}\oplus V_2\oplus e^{i\phi_2},
 \qquad V_2\sim\mathrm{Haar}[U(2)].
 \label{eq:sm_typeII_u1_gate}
\end{equation}
The phases $\phi_0$ and $\phi_2$ are random and the $U(2)$ block is
independent for every gate and every layer. One circuit period comprises
an even-bond and an odd-bond layer (each bond acts on two neighboring
sites), $U(t)=U_{\mathrm{odd}}(t)U_{\mathrm{even}}(t)$. Hence $t$ counts pairs of sublayers. Consequently, the circuit conserves $Q=\sum_j n_j$ exactly, even for a single circuit realization.
\vspace{0.3cm}
\paragraph{The charge-diffusion eigenvalue.}
Each gate is unitary and conserves charge in every realization.
An independent Haar average is useful because it turns the one-charge
block into a completely mixed channel, so the occupations obey
a closed linear update. After an active bond is updated,
\begin{equation}
 \mathbb E\langle n_j'\rangle
 =\mathbb E\langle n_{j+1}'\rangle
 =\frac12\bigl(\langle n_j\rangle+\langle n_{j+1}\rangle\bigr).
 \label{eq:sm_typeII_bond_average}
\end{equation}
Let $M_{\mathrm e}$ and $M_{\mathrm o}$ replace the occupations
on all even and odd bonds, respectively, by their pair averages.
The two-layer averaged transfer matrix is $M_{\mathrm o}M_{\mathrm e}$.

For even $N$, one can equally view one period as a map on the
$L=N/2$ densities $u_j$ immediately after an even-bond average:
\begin{equation}
 u_j(t+1)=\tfrac14u_{j-1}(t)+\tfrac12u_j(t)
 +\tfrac14u_{j+1}(t),
 \label{eq:sm_typeII_diffusion_lattice}
\end{equation}
with reflecting ends $u_0=u_1$ and $u_{L+1}=u_L$
for sites labeled $j=1,\ldots,L$. The eigenmodes are discrete
cosines with eigenvalues
$\Lambda_k=\cos^2(k\pi/N)$, $k=0,\ldots,L-1$.
The uniform $k=0$ mode represents conserved total charge; the first
nonuniform mode has
\begin{equation}
 \Lambda_1=\cos^2\left(\frac{\pi}{N}\right),\qquad
 \gamma_Q=-\log\Lambda_1
 =\frac{\pi^2}{N^2}+O(N^{-4}),\qquad
 \tau_Q = 1/\gamma_Q.
 \label{eq:sm_typeII_gamma}
\end{equation}
This equation concerns the \emph{averaged density channel}; it does not
identify the slowest relaxation rate of every entanglement observable
of each pure-state realization.

For odd $N=2L+1$, the even-bond layer leaves the last site unpaired. Denoting its density by $u_{L+1}$, the bulk recursion remains unchanged, while the last site obeys
\begin{equation}
u_{L+1}(t+1)=\frac12[u_L(t)+u_{L+1}(t)].
\end{equation}
The cosine ansatz $u_j\propto\cos[q(j-\frac12)]$ then gives $\sin[q(L+\frac12)]=0$, yielding $q=2\pi k/N$ and $\Lambda_k=\cos^2(k\pi/N)$, $k=0,\ldots,L$. Thus the slowest nonuniform eigenvalue $\Lambda_1$ and its decay rate $\gamma_Q$ are identical for even and odd $N$.

\vspace{0.3cm}
\paragraph{Initial preparation and operational times.} 
We initialize the open chain in the charge-modulated product state
\begin{equation}
\begin{split}
|\psi_0\rangle
&=\bigotimes_{j=1}^{N}
\left(\sqrt{1-p_j}|0\rangle_j
+\sqrt{p_j}|1\rangle_j\right),~~~~~~~
p_j=\frac12+0.4\cos\left[
\frac{\pi(j-\frac12)}{N}\right],
\end{split}
\label{eq:sm_typeII_hydro_state}
\end{equation}
which overlaps with the slowest long-wavelength charge mode. We consider $N=6,9,12,15,18$ and divide each chain into three equal consecutive regions $A|B|C$. For each pure-state realization, we evaluate the spatially averaged ordinary entropies $S_{n,\mathrm{av}}=\frac13\sum_{X=A,B,C}S_n(X)$, with $n=1,2$, and the genuine multi-entropy $G=GM_2^{(3)}$.

To monitor charge diffusion, we define
\begin{equation}
Q_1(t)=\sum_{j=1}^{N}
\cos\left[\frac{\pi(j-\frac12)}{N}\right]
\left(\langle n_j(t)\rangle-\frac12\right),
\label{eq:sm_typeII_Q1}
\end{equation}
where $n_j=(\mathbb I-Z_j)/2$. In each circuit realization $r$, the local occupation $\langle n_j(t)\rangle_r= \langle\psi_r(t)|n_j|\psi_r(t)\rangle$ is evaluated directly from the evolved pure state $|\psi_r(t)\rangle =U_{\mathrm{odd}}^{(r)}(t-1) U_{\mathrm{even}}^{(r)}(t-1)|\psi_r(t-1)\rangle.$ The ensemble-averaged charge modulation is then $\overline Q_1(t)=R^{-1}\sum_{r=1}^{R}Q_{1,r}(t)$, where $R$ denotes the number of independent circuits.


The simulations cover $0\leq t\leq100$ in integer two-layer circuit periods. The realization counts for $N=6,9,12,15,18$ are $R=24,16,8,4,2$, respectively. Each nonlinear entanglement observable is evaluated separately in each realization before ensemble averaging. For the resulting ensemble-averaged observable $\mathcal O(t)$, we define
\begin{equation}
\begin{split}
\mathcal O_{\mathrm{ref}}
&=\frac1{21}\sum_{u=80}^{100}\mathcal O(u),~~~~~~~~~~
t_{90}[\mathcal O]
=\min\left\{t:
\mathcal O(t)\geq
\mathcal O(0)+0.9[
\mathcal O_{\mathrm{ref}}-\mathcal O(0)]
\right\}.
\end{split}
\label{eq:sm_typeII_t90}
\end{equation}
Thus $t_{90}^{S_1}$, $t_{90}^{S_2}$, and $t_{90}^{G}$ measure the first $90\%$ growth crossings of the corresponding ensemble-averaged observables toward their late-window references.

For the charge modulation, we instead define the first $90\%$ amplitude-decay time,
\begin{equation}
t_{90}^{Q_1}
=\min\left\{t:
|\overline Q_1(t)|
\leq0.1|Q_1(0)|
\right\}.
\label{eq:sm_typeII_charge_t90}
\end{equation}
This numerical crossing time is distinct from the analytically determined spectral decay time of the slowest nonuniform charge mode,
\begin{equation}
\tau_Q=\frac1{\gamma_Q}
=\left[-2\ln\cos\left(\frac{\pi}{N}\right)\right]^{-1}
\simeq\frac{N^2}{\pi^2}.
\label{eq:sm_typeII_tauQ}
\end{equation}
The latter characterizes a $1/e$ amplitude decay, whereas $t_{90}^{Q_1}$ measures a $90\%$ reduction of the numerically observed charge modulation. The operational crossing times and spectral decay times for all five system sizes are summarized in Table~\ref{tab:sm_typeII_u1_times}.
\begin{table}[h!]
\centering
\begingroup
\setlength{\tabcolsep}{10pt}
\renewcommand{\arraystretch}{1.3}
\begin{tabular}{c|rrrrr}
\hline\hline
$N$ & $t_{90}^{S_1}$ & $t_{90}^{S_2}$
& $t_{90}^{G}$ & $t_{90}^{Q_1}$ & $\tau_Q$\\
\hline
6  & 6  & 7  & 10 & 10 & 3.476\\
9  & 9  & 11 & 14 & 18 & 8.038\\
12 & 12 & 16 & 22 & 33 & 14.422\\
15 & 15 & 19 & 26 & 50 & 22.630\\
18 & 20 & 28 & 36 & 76 & 32.661\\
\hline\hline
\end{tabular}
\endgroup
\caption{\small Operational $90\%$ crossing times for the ensemble-averaged observables in the $U(1)$ circuit. The charge crossing $t_{90}^{Q_1}$ measures a $90\%$ amplitude reduction, while $\tau_Q$ is obtained independently from the averaged density-transfer spectrum. All times are expressed in complete two-layer circuit periods.}
\label{tab:sm_typeII_u1_times}
\end{table}

\begin{figure*}[h!]
\centering
\includegraphics[width=0.73\textwidth]{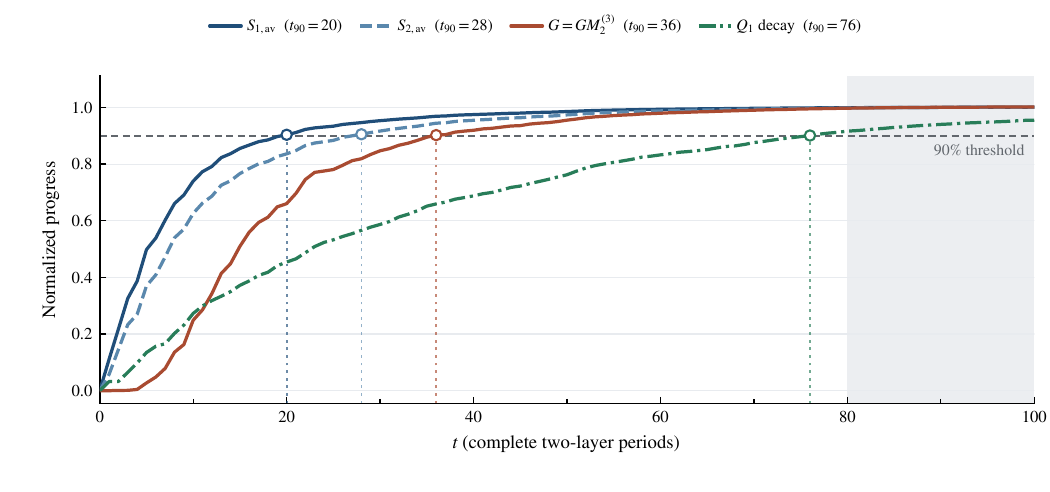}
\caption{\small Normalized entanglement growth and charge relaxation in the $N=18$ $U(1)$-conserving random circuit. Solid dark-blue, dashed light-blue, solid orange, and dash-dotted green curves denote $S_{1,\mathrm{av}}$, $S_{2,\mathrm{av}}$, $G=GM_2^{(3)}$, and the charge modulation $Q_1$, respectively. Entanglement observables are normalized as $F_O(t)=[\overline O(t)-\overline O(0)]/[\overline O_{\mathrm{ref}}-\overline O(0)]$, where $\overline O_{\mathrm{ref}}$ is averaged over $80\leq t\leq100$; charge relaxation is shown as $F_Q(t)=1-|\overline Q_1(t)|/|Q_1(0)|$. The horizontal dashed line marks the $90\%$ threshold, with first crossings at $t_{90}=(20,28,36,76)$, respectively (open circles). The shaded region indicates the reference window. Results are averaged over two independent circuit realizations, with time measured in complete two-layer periods.}
\label{fig:sm_u1_n18_one_axes}
\end{figure*}

For $N=18$, the normalized evolution of the entanglement observables and charge modulation is shown in Fig.~\ref{fig:sm_u1_n18_one_axes}. The spectral prediction for a $90\%$ decay of the slowest charge eigenmode is $t_{Q,90}^{\mathrm{spec}}=\tau_Q\ln10\simeq75.2$ periods, close to the measured $t_{90}^{Q_1}=76$. At the GM crossing time $t_{90}^{G}=36$, the ensemble-averaged charge modulation still retains approximately $34\%$ of its initial amplitude. The slow diffusive mode therefore continues to relax after the dominant growth of the entanglement observables has largely subsided.

Across the five system sizes $N=6,9,12,15,18$, unconstrained finite-size fits give $t_{90}^{S_1}\propto N^{1.067}$, $t_{90}^{G}\propto N^{1.159}$, and $\tau_Q\propto N^{2.039}$. The approximately linear scaling of the entanglement growth times is consistent with propagation-limited dynamics, in contrast to the diffusive $N^2$ charge timescale. However, these operational results do not exclude weak late-time hydrodynamic contributions to entanglement, and the limited number of realizations at larger $N$ prevents a definitive asymptotic scaling conclusion.

\subsection{Five-qubit static code: perturbative logical dynamics and independent clocks}
\label{sec:sm_typeII_static_code}

\paragraph{Code space, microscopic Hamiltonian, and initial state.}
We use the $[[5,1,3]]$ stabilizer code to separate local
syndrome dressing from a slow coherent logical process under
a \emph{time-independent} Hamiltonian.  Label the physical qubits
$i=0,\ldots,4$ and take the commuting stabilizer generators
\begin{equation}
\begin{aligned}
S_1=XZZXI, ~~~~~S_2=IXZZX,~~~~~
S_3=XIXZZ, ~~~~~S_4=ZXIXZ.
\end{aligned}
\label{eq:sm_static_stabilizers}
\end{equation}
where each Pauli string is ordered from site $0$ to site $4$.
The projector onto their common $+1$ eigenspace and its complement are
\begin{equation}
P=\prod_{a=1}^{4}\frac{\mathbb I+S_a}{2},
\qquad Q=\mathbb I-P.
\label{eq:sm_static_projector}
\end{equation}
The code space has dimension two; we choose
$\overline X=X^{\otimes5}$ and $\overline Z=Z^{\otimes5}$ as
physical representatives of the logical Pauli operators and define
\begin{equation}
|\overline 0\rangle=\frac{P|00000\rangle}{\|P|00000\rangle\|},
\qquad |\overline 1\rangle=\overline X|\overline 0\rangle.
\label{eq:sm_static_logical_basis}
\end{equation}
To distinguish the abstract logical qubit from its
five-qubit physical encoding, introduce the isometry
\begin{equation}
\begin{split}
W:\mathcal H_L&\rightarrow\mathcal H_{\mathrm{phys}},
~~~~~~
W|0\rangle_L=|\overline 0\rangle,\qquad
W|1\rangle_L=|\overline 1\rangle,
\end{split}
\end{equation}
where $\mathcal H_L\simeq\mathbb C^2$ and
$\mathcal H_{\mathrm{phys}}=(\mathbb C^2)^{\otimes5}$.
The isometry satisfies $W^\dagger W=\mathbb I_L$ and $WW^\dagger=P$.
Thus $|0\rangle_L$ and $|\overline 0\rangle$ represent the same logical information in different
Hilbert spaces, but are not identical state vectors.
The initial state is $|\psi(0)\rangle=|\overline 0\rangle$.
For the tripartition $A=(0,1)$, $B=(2,3)$, and $C=(4)$,
all regions are correctable erasures.  In particular, the
distance-three code has vanishing code-space matrix elements
for every nonidentity Pauli operator supported on at most two
sites.  Consequently, \emph{every} state within the ideal code
space, including arbitrary logical superpositions, satisfies
\begin{equation}
\rho_A=\frac{\mathbb I_4}{4},\qquad
\rho_B=\frac{\mathbb I_4}{4},\qquad
\rho_C=\frac{\mathbb I_2}{2}.
\label{eq:sm_static_ideal_marginals}
\end{equation}
Thus logical rotations leave all three complete single-party
Schmidt spectra unchanged, although their multipartite
local-unitary invariants can vary.

The physical Hamiltonian is
\begin{equation}
H(g) =H_0+gV,\qquad H_0=-\kappa\sum_{a=1}^{4}S_a,\qquad
V=\sum_{i=0}^{4}\sum_{\alpha=x,y,z}h_i^\alpha\sigma_i^\alpha.
\label{eq:sm_static_H}
\end{equation}
The four stabilizers commute and square to the identity;
$H_0$ therefore has code energy $E_0=-4\kappa$ and
syndrome energies $E_s=E_0+2\kappa|s|$, where
$|s|$ counts violated stabilizers.  The minimum gap is
$\Delta=2\kappa$.  We set $\kappa=\hbar=1$.
Unlike a circuit model, this is a static five-qubit Hamiltonian
with four-body stabilizer interactions; it is not a
thermalizing nearest-neighbor chain.

The same normalized field realization is used at every $g$:
$h_i^\alpha=r_i^\alpha/\mathcal N$, where
$\mathcal N^2=\sum_{i,\alpha}(r_i^\alpha)^2$.
The rows below enumerate sites $i=0,\ldots,4$ and the columns correspond to $x$, $y$, and $z$:
\begin{table}[H]
  \centering
\setlength{\tabcolsep}{10pt}
\renewcommand{\arraystretch}{1.3}
  \begin{tabular}{c|ccc}
\hline \hline
 $i$&$r_i^x$&$r_i^y$&$r_i^z$\\ \hline
0& 0.65622409&-0.23312355&-0.90402765\\
1&-0.45172139&-0.05826052&-0.21048139\\
2&-0.56589406&-1.41514039&-0.83941102\\
3& 1.23284335&-1.45794580&-0.32780708\\
4& 0.48432356&-1.14060282&-0.91677836 \\
\hline \hline
\end{tabular}
\label{eq:sm_static_fields}
\end{table}
Here $\mathcal N\simeq3.27268294$, so that $\sum_{i,\alpha}(h_i^\alpha)^2=1$; the disorder pattern is \emph{not} resampled as the perturbation strength changes.

\vspace{0.3cm}
\paragraph{Fast syndrome dressing and suppressed entropy response.}
A single-site Pauli error is detectable by the code, so
$P\sigma_i^\alpha P=0$ and hence $PVP=0$.
Nevertheless, $QVP$ is generally nonzero.  To first order
in $g$, the excited-syndrome component of the state is
\begin{equation}
\begin{aligned}
    &Q|\psi(t)\rangle= Q e^{-iHt} | \psi(0) \rangle =
g\sum_{n\in\operatorname{ran}Q}
\frac{e^{-iE_nt}-e^{-iE_0t}}{E_n-E_0}
|n\rangle\langle n|V|\overline 0\rangle+O(g^2),\\
&e^{-iE_nt}-e^{-iE_0t}
=-2ie^{-i(E_n+E_0)t/2}
\sin\left(\frac{E_n-E_0}{2}t\right).
\end{aligned}
\label{eq:sm_static_dyson}
\end{equation}
where $H_0|n\rangle=E_n|n\rangle, Q=\sum_{n\in\operatorname{ran}Q}|n\rangle\langle n|$. Because
$E_n-E_0\geq\Delta$, the off-resonant amplitude is
$O(g/\Delta)$ after microscopic times
$t_{\mathrm{local}}=O(\Delta^{-1})$, with a leakage
probability of order $g^2/\Delta^2$.  The first-order Dyson
expansion alone is not uniform on the much longer logical
timescale; for sufficiently weak $g$, the isolated dressed
low-energy doublet supplies the corresponding control
of syndrome admixture.

In the full physical evolution the code-space identities
in Eq.~\eqref{eq:sm_static_ideal_marginals} are therefore
only approximate: the reduced matrices can have corrections
$\rho_X=\mathbb I_{d_X}/d_X+\delta\rho_X$ with
$\delta\rho_X=O(g/\Delta)$.
Since $\operatorname{Tr}\delta\rho_X=0$, the entropy
is stationary to first order around the maximally mixed state,
\begin{equation}
S_1(X)=\ln d_X-\frac{d_X}{2}
\operatorname{Tr}(\delta\rho_X)^2
+O(\|\delta\rho_X\|^3),
\label{eq:sm_static_entropy_expansion}
\end{equation}
and similarly
$S_2(X)=\ln d_X-d_X\operatorname{Tr}(\delta\rho_X)^2
+O(\|\delta\rho_X\|^4)$.
Therefore the local entropy drift is $O(g^2/\Delta^2)$,
even though the reduced matrices may change at first order.
This is suppressed \emph{dressing}, not an exact marginal
plateau of the perturbed physical state.
\vspace{0.3cm}
\paragraph{Third-order effective logical Hamiltonian.}
Introduce the reduced resolvent in the excited space,
\begin{equation}
R_Q=Q(E_0-H_0)^{-1}Q
=\sum_{n\in\operatorname{ran}Q}
\frac{|n\rangle\langle n|}{E_0-E_n}.
\label{eq:sm_static_resolvent}
\end{equation}

The low-energy expansion below is a form of degenerate
Schrieffer--Wolff perturbation theory
\cite{BravyiDiVincenzoLoss2011}. 
Writing $V=\sum_\mu h_\mu E_\mu$ as a sum of single-site
Pauli errors, the code distance implies
\begin{equation}
PVP=0,\qquad
PVR_QVP=-\sum_\mu\frac{h_\mu^2}{\Delta_\mu}P
\equiv c_2P,
\label{eq:sm_static_second_order}
\end{equation}
where $\Delta_\mu>0$ is the excitation energy of the
syndrome created by $E_\mu$.  In the second expression,
products of distinct one-site Pauli errors either have
nontrivial weight at most two and vanish between code
projectors, or, for identical errors, reduce to the identity.
Thus neither first nor second order generates a
nontrivial logical rotation.

At third order, an ordered error sequence can return to the
code space after two nonzero intermediate syndromes.
Consider three successive single-site Pauli errors
$E_c$, $E_b$, and $E_a$, acting from right to left.
Starting from the code space with syndrome zero,
the successive syndromes are
\begin{equation}
0
\xrightarrow{E_c}s_c
\xrightarrow{E_b}s_c\oplus s_b
\xrightarrow{E_a}s_c\oplus s_b\oplus s_a.
\end{equation}
A contribution to $PVR_QVR_QVP$ requires both
intermediate syndromes to be nonzero,
$s_c\neq0$ and $s_c\oplus s_b\neq0$,
while the final syndrome must vanish,
$s_a\oplus s_b\oplus s_c=0$.
The two virtual excitations contribute the energy
denominator
$[\Delta(s_c)\Delta(s_b\oplus s_c)]^{-1}$,
where $\Delta(s)$ is the excitation energy of
syndrome sector $s$ relative to the code space.
Weight-three Pauli products can act nontrivially on the
encoded qubit.  If $W=(|\overline 0\rangle,|\overline 1\rangle)$ is the
encoding isometry, the effective two-dimensional Hamiltonian is
\begin{equation}
\begin{split}
H_{\mathrm{eff}}^{(L)}
={}&(E_0+g^2c_2)\mathbb I_2
+g^3 W^\dagger VR_QVR_QVW+O(g^4/\Delta^3)
=E_{\mathrm{shift}}(g)\mathbb I_2
+g^3\boldsymbol h_3\cdot\boldsymbol\sigma_L
+O(g^4/\Delta^3),
\end{split}
\label{eq:sm_static_Heff}
\end{equation}
where a possible scalar term at cubic order is absorbed
into $E_{\mathrm{shift}}(g)$, and
$\boldsymbol h_3=O(\Delta^{-2})$.
For the fixed fields above, explicit evaluation gives
$2|\boldsymbol h_3|\simeq0.07385436$ at $\kappa=1$.
Consequently the low-energy logical-doublet splitting and
coherent timescale obey
\begin{equation}
\Delta E_L(g)=2|\boldsymbol h_3|g^3
+O(g^4/\Delta^3),\qquad
\tau_L\sim\frac{\Delta^2}{g^3},
\label{eq:sm_static_gap_scale}
\end{equation}
whereas $t_{\mathrm{local}}=O(\Delta^{-1})$.
The distance-three condition permits, but does not
by itself guarantee, a nonzero cubic logical term:
its coefficient depends on the chosen local fields $h_i^\alpha=r_i^\alpha/\mathcal N$.

\vspace{0.3cm}
\paragraph{Multipartite response to logical rotation.}
We first examine the response of $G$ strictly \emph{within}
the ideal code space. For a pure logical state
$\rho_L=\frac12(\mathbb I_L+
\mathbf r\cdot\boldsymbol{\sigma}_L)$,
where $\mathbf r=(r_x,r_y,r_z)$ and $|\mathbf r|=1$,
the physical encoded state is
$\rho_{\mathrm{code}}=W\rho_LW^\dagger$.

General methods for calculating multi-invariants of stabilizer states
have been developed in Refs.~\cite{AkellaGaddePandey2026,CzechEtAl2026}.
Here we evaluate the restricted permutation operator for the stated
encoding and tripartition explicitly.

Restricting the four-replica permutation operator
to this subspace gives
\begin{equation}
K_{\mathrm{code}}
=(W^\dagger)^{\otimes4}\mathcal P_{\mathrm{code}}
W^{\otimes4}
=\frac1{64}\sum_{\mu=0,x,y,z}
(\sigma_\mu^{(L)})^{\otimes4},
\label{eq:sm_typeII_Kcode}
\end{equation}
where
$\mathcal P_{\mathrm{code}}=
P_A((13)(24))P_B((12)(34))$,
$\sigma_0^{(L)}=\mathbb I_L$, and
$(\sigma_x^{(L)},\sigma_y^{(L)},\sigma_z^{(L)})
=(X_L,Y_L,Z_L)$.
The kernel follows from the stabilizer-projector
contraction.

Using
$\operatorname{Tr}(\rho_L\sigma_i^{(L)})=r_i$,
we obtain
\begin{equation}
Z_2^{(3)}
=\operatorname{Tr}
\left(\rho_L^{\otimes4}K_{\mathrm{code}}\right)
=\frac{1+r_x^4+r_y^4+r_z^4}{64}.
\end{equation}
Every state strictly within the code space satisfies
$\rho_A=\mathbb I_4/4$,
$\rho_B=\mathbb I_4/4$, and
$\rho_C=\mathbb I_2/2$.
Consequently,
$\sum_X S_2(X)=\ln32$, yielding
\begin{equation}
G_{\mathrm{code}}(\mathbf r)
=\frac12\ln
\frac{2}{1+r_x^4+r_y^4+r_z^4}.
\label{eq:sm_static_logical_G}
\end{equation}
Thus the single-party entropies are insensitive
to logical rotations within the ideal code space,
whereas $G_{\mathrm{code}}$ depends explicitly
on the logical Bloch-vector orientation.

The third-order effective Hamiltonian
$H_{\mathrm{eff}}^{(L)}
=E_{\mathrm{shift}}\mathbb I_L+
g^3\mathbf h_3\cdot\boldsymbol{\sigma}_L$
generates coherent precession of the logical
Bloch vector,
\begin{equation}
\dot{\mathbf r}
=\Delta E_L\,
\hat{\mathbf h}_3\times\mathbf r,
\qquad
\Delta E_L=2g^3|\mathbf h_3|,
\label{eq:sm_static_bloch_motion}
\end{equation}
at leading order in $g$.
Writing
$\hat{\mathbf h}_3=(n_x,n_y,n_z)$
and $\theta=\Delta E_Lt$, the initial
logical state $|0\rangle_L$, with
$\mathbf r(0)=(0,0,1)$, evolves as
\begin{equation}
\begin{aligned}
r_x(t)
&=n_y\sin\theta
+n_xn_z(1-\cos\theta),\\
r_y(t)
&=-n_x\sin\theta
+n_yn_z(1-\cos\theta),\\
r_z(t)
&=n_z^2+(1-n_z^2)\cos\theta.
\end{aligned}
\label{eq:sm_static_bloch_components}
\end{equation}
Substituting these expressions into
Eq.~\eqref{eq:sm_static_logical_G}
gives
\begin{equation}
G_{\mathrm{code}}(t)
=\frac12\ln
\frac{2}{
1+r_x^4(t)+r_y^4(t)+r_z^4(t)
}.
\label{eq:sm_static_G_oscillation}
\end{equation}
For our fixed perturbation,
$(h_{3x},h_{3y})\neq(0,0)$,
so the logical rotation produces a
nonconstant, periodic $G_{\mathrm{code}}(t)$.
Although $G_{\mathrm{code}}$ need not
oscillate sinusoidally, its time
dependence is governed by the
dimensionless phase $\Delta E_Lt$.
Consequently, the first half-rise occurs
at a finite angle $\theta_{1/2}$,
implying
\begin{equation}
t_{1/2}^{G}
\simeq\frac{\theta_{1/2}}{\Delta E_L}
\propto g^{-3}
\end{equation}
in the weak-coupling regime.

Importantly, these exact code-space identities do \emph{not} hold for the full physical evolution at finite $g$. Since $QVP\neq0$, the perturbation generates excited-syndrome admixture and corrections to the ideal marginals. The expression $G_{\mathrm{code}}(\mathbf r(t))$ describes the leading coherent logical contribution, while the full $G(t)$ also contains perturbative dressing corrections.

\vspace{0.3cm}
\paragraph{Independent dynamical and amplitude diagnostics.}
We evolve the full $32$-dimensional Hilbert space with
$|\psi(t)\rangle=e^{-iH(g)t}|\overline 0\rangle$ and compare
it with the effective logical evolution.  For each pure state
we evaluate $S_n(X)$ and the same four-copy genuine
multi-entropy $G=GM_2^{(3)}$ as in the main text.
To display the coherent oscillations, we may use the phase
$x=\Delta E_L(g)t$; this rescaling is used only for
visualization, not to extract the observable clock.

Define $S_{\mathrm{av}}=(S_1(A)+S_1(B)+S_1(C))/3$.
On a prescribed physical-time sample $\mathcal T$ containing
both microscopic times and the logical oscillation,
the measured response amplitudes are
\begin{equation}
\begin{split}
\Delta G=\max_{t\in\mathcal T}G(t)
-\min_{t\in\mathcal T}G(t),~~~~~~~~~~~~~~~~
\Delta S_{\mathrm{av}}
=\max_{t\in\mathcal T}
|S_{\mathrm{av}}(t)-S_{\mathrm{av}}(0)|.
\end{split}
\label{eq:sm_static_amplitudes}
\end{equation}
The fast local timescale $t_{1/2}^{S_1}$
is defined by the first rising crossing of half
the maximum entropy-deficit drift in the early-time window.
To determine the slow clock independently of
the spectral splitting, we use for every $g$
the same absolute-time sampling window:
$t=0$ and $2400$ logarithmically spaced times
over $0.1\leq t\leq2\times10^6$ (with $\hbar=1$). Let
$G_{\max}^{\mathrm{abs}}(g)=\max_{t\in\mathcal T_{\mathrm{abs}}}G(t;g)$.
The first rising half-height time is defined by
\begin{equation}
G(t_{1/2}^{G};g)=G(0;g)
+\frac12\bigl[G_{\max}^{\mathrm{abs}}(g)-G(0;g)\bigr],
\label{eq:sm_static_halfclock}
\end{equation}
with the first upward crossing bracketed by the sampled trajectory and refined using exact spectral evolution. We independently check the first crossing of the fixed absolute increment $G(t)-G(0)=0.1$. Neither the fitted $\Delta E_L(g)$ nor the scaled phase $x$ is used to select the sampling times or determine these crossings.  Doubling the number of time samples changes the reported half-height crossings by at most $0.94\%$ over the seven couplings examined.

Over $0.035\leq g\leq0.15$, independent numerical fits yield
\begin{equation}
\begin{split}
\Delta E_L(g)&\simeq0.0738770\,g^{3.00010},\\
t_{1/2}^{G}(g)&\simeq7.06608\,g^{-3.00427},\\
t_{0.1}^{G}(g)&\simeq7.06897\,g^{-3.00326}.
\end{split}
\label{eq:sm_static_numeric_clocks}
\end{equation}
The first coefficient agrees with the independently calculated $2|\boldsymbol h_3|$.  The two separately measured GM crossing times reproduce the cubic coherent scale without imposing a $g^{-3}$ sampling grid. Meanwhile, the fast entropy-dressing markers remain $O(g^0)$. The mean-entropy drift scales approximately as $\Delta S_{\mathrm{av}}\propto g^{1.9544}$, while the sampled GM excursion stays of order $0.2$. In particular, for $g=0.08$,
\begin{equation}
\Delta S_{\mathrm{av}}=0.00130996,
\qquad \Delta G=0.200482.
\label{eq:sm_static_numeric_amplitudes}
\end{equation}
These are three complementary checks of the same mechanism: a coherent $O(1)$ multipartite response, parametrically slow logical motion on $g^{-3}$ time, and quadratically suppressed local entropy drift. The static-code clock characterizes \emph{coherent logical oscillations}, not irreversible thermalization, entropy saturation, or a distinct asymptotic relaxation gap. Moreover, an initial logical eigenstate of the effective Hamiltonian, or a different partition, need not display the same GM response.

\subsection{Logical information coupled to a locally scrambling bath}
\label{sec:sm_typeII_local_bath}
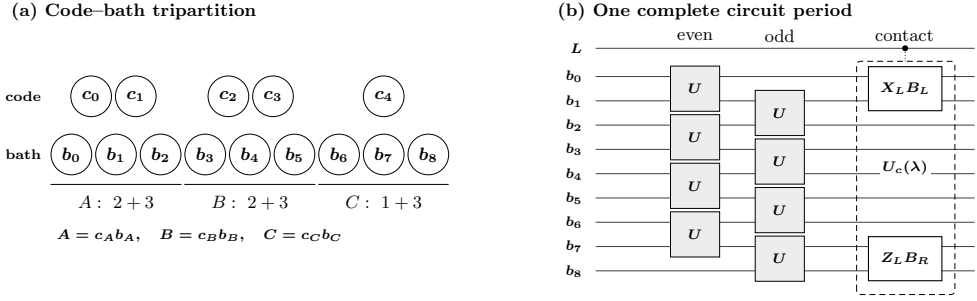
\begin{figure*}[h!]
\centering
\resizebox{0.80\textwidth}{!}{%
\begin{minipage}{\textwidth}
\centering
\begin{minipage}[t]{0.47\textwidth}
\centering
\resizebox{!}{4.95cm}{%
\begin{tikzpicture}[
 x=1cm,y=1cm,
 every node/.style={text=black},
 site/.style={circle,draw=black,very thin,minimum size=7.2mm,inner sep=0pt,font=\small\boldmath},
 thinrule/.style={black,line width=0.48pt},
 font=\small
]
\path[use as bounding box] (-0.10,-0.18) rectangle (8.12,5.12);
\node[anchor=west,font=\bfseries\small] at (0,4.93) {(a) Code--bath tripartition};
\begin{scope}[yshift=-5mm]
\node[anchor=east,font=\scriptsize\bfseries] at (0.76,3.93) {code};
\node[anchor=east,font=\scriptsize\bfseries] at (0.76,2.90) {bath};
\foreach \i/\xx in {0/1.53,1/2.32,2/3.96,3/4.75,4/6.71}{
  \node[site] at (\xx,3.93) {$c_{\i}$};
}
\foreach \i in {0,...,8}{
  \pgfmathsetmacro{\xx}{1.18+0.79*\i}
  \node[site] at (\xx,2.90) {$b_{\i}$};
}
\draw[thinrule] (0.81,2.37)--(3.13,2.37);
\draw[thinrule] (3.18,2.37)--(5.50,2.37);
\draw[thinrule] (5.55,2.37)--(7.86,2.37);
\node[font=\small] at (1.97,2.04) {$A:\ 2+3$};
\node[font=\small] at (4.34,2.04) {$B:\ 2+3$};
\node[font=\small] at (6.71,2.04) {$C:\ 1+3$};
\node[anchor=west,font=\scriptsize\boldmath,text=black] at (0.81,1.47)
 {$A=c_Ab_A,\quad B=c_Bb_B,\quad C=c_Cb_C$};
\end{scope}
\end{tikzpicture}%
}
\end{minipage}\hfill
\begin{minipage}[t]{0.51\textwidth}
\centering
\resizebox{!}{4.95cm}{%
\begin{tikzpicture}[
 x=1cm,y=1cm,
 every node/.style={text=black},
 wire/.style={black!75,line width=0.45pt},
 twogate/.style={draw=black,line width=0.50pt,fill=black!7,
    minimum width=8.7mm,minimum height=8.0mm,inner sep=0pt,font=\scriptsize\boldmath},
 cgate/.style={draw=black,line width=0.55pt,fill=white,
    minimum width=13mm,minimum height=7.8mm,inner sep=1pt,font=\scriptsize\boldmath},
 font=\small
]
\path[use as bounding box] (0,-0.18) rectangle (8.35,5.12);
\node[anchor=west,font=\bfseries\small] at (0.05,4.93) {(b) One complete circuit period};
\node[font=\small] at (2.60,4.50) {even};
\node[font=\small] at (4.10,4.50) {odd};
\node[font=\small] at (6.31,4.50) {contact};
\node[anchor=east,font=\scriptsize\boldmath] at (0.72,4.28) {$L$};
\draw[wire] (0.84,4.28)--(7.55,4.28);
\foreach \j in {0,...,8}{
  \pgfmathsetmacro{\yy}{3.78-0.43*\j}
  \node[anchor=east,font=\scriptsize\boldmath] at (0.72,\yy) {$b_{\j}$};
  \draw[wire] (0.84,\yy)--(7.55,\yy);
}
\foreach \j in {0,2,4,6}{
  \pgfmathsetmacro{\yc}{3.78-0.43*\j-0.215}
  \node[twogate] at (2.60,\yc) {$U$};
}
\foreach \j in {1,3,5,7}{
  \pgfmathsetmacro{\yc}{3.78-0.43*\j-0.215}
  \node[twogate] at (4.10,\yc) {$U$};
}
\draw[densely dashed,black,line width=.50pt,rounded corners=2pt]
   (5.46,-0.08) rectangle (7.20,4.05);
\fill (6.32,4.28) circle (1.5pt);
\draw[densely dotted,black] (6.32,4.21)--(6.32,4.05);
\node[cgate] at (6.32,3.565) {$X_L B_L$};
\node[cgate] at (6.32,0.555) {$Z_L B_R$};
\node[fill=white,inner sep=1.5pt,font=\scriptsize\boldmath] at (6.32,2.17)
 {$U_c(\lambda)$};
\end{tikzpicture}%
}
\end{minipage}
\end{minipage}%
}%
\caption{\small Code-preserving coupling to a locally scrambling nine-qubit bath.
(a) The $[[5,1,3]]$ code is partitioned into $2|2|1$ physical qubits,
and the bath into three consecutive $3$-qubit blocks, defining the
pure tripartition $A|B|C$. 
(b) Each period applies independent nearest-neighbor Haar $U(4)$ gates
on the even and odd bonds, followed by the single joint contact
$U_c(\lambda)=\exp[-i\lambda(X_L\otimes B_L+
Z_L \otimes B_R)]$. The dashed enclosure denotes one joint
exponential, not two sequential contact gates. All Haar gates are
resampled independently at each period.}
\label{fig:m2-supplement-circuit}
\end{figure*}

\vspace{0.3cm}
\paragraph{Circuit, encoding, and tripartition.}
We retain the $[[5,1,3]]$ stabilizer code described above, but couple its encoded logical qubit to an open chain of $N_b=9$ bath qubits. The numerical evolution is carried out in the \emph{reduced logical--bath space} $\mathcal H_L\otimes\mathcal H_b\simeq \mathbb C^2\otimes(\mathbb C^2)^{\otimes9}$; the corresponding physical state is obtained by applying the five-qubit encoding isometry $W$ to the logical factor. The code shares are $c_A=(0,1)$, $c_B=(2,3)$, and $c_C=(4)$, while the consecutive bath shares are $b_A=(0,1,2)$, $b_B=(3,4,5)$, and $b_C=(6,7,8)$.  The three parties $A=c_Ab_A$, $B=c_Bb_B$, and $C=c_Cb_C$ exhaust the 14 physical qubits, with dimensions $(d_A,d_B,d_C)=(32,32,16)$. The code--bath tripartition is illustrated in Fig.~\ref{fig:m2-supplement-circuit}(a).

The initial logical--bath state is
\begin{equation}
 |\Psi(0)\rangle_{Lb}
 =|0\rangle_L \otimes
 \bigotimes_{j=0}^{8}|\varphi_j\rangle,
 \label{eq:sm_bath_initial}
\end{equation}
Here $W|0\rangle_L=|\overline 0\rangle$ is the physical encoded initial state, thus the corresponding physical state is
$(W\otimes\mathbb I_b)|\Psi(0)\rangle_{Lb}=|\overline 0\rangle\otimes \bigotimes_{j=0}^{8}|\varphi_j\rangle$. 
The nine single-qubit pure states are sampled independently and uniformly on the Bloch sphere:
$|\varphi_j\rangle=\cos(\theta_j/2)|0\rangle+
 e^{i\phi_j}\sin(\theta_j/2)|1\rangle$, with
$\cos\theta_j$ uniform on $[-1,1]$ and $\phi_j$
uniform on $[0,2\pi)$.
Each complete period applies independent Haar $U(4)$ gates
on the even bath bonds $(0,1),(2,3),(4,5),(6,7)$,
then on the odd bonds $(1,2),(3,4),(5,6),(7,8)$, followed by one joint logical contact,
\begin{equation}
\begin{split}
 U_t&=U_c(\lambda)U_{\mathrm odd}(t)U_{\mathrm even}(t),\\
 U_c(\lambda) =e^{-i\lambda K},\qquad
 K &=X_L\otimes B_L+Z_L\otimes B_R,~~~~~
 B_L=Y_0X_1,\qquad B_R=X_7Y_8.
\end{split}
\label{eq:sm_bath_period}
\end{equation}
The complete circuit period is illustrated in Fig.~\ref{fig:m2-supplement-circuit}(b). Here $X_L,Z_L$ act on the logical qubit and are represented
in the physical code by $\overline X=X^{\otimes5}$ and
$\overline Z=Z^{\otimes5}$.  Every bath gate is freshly
and independently sampled in space and time.  The
logical operators preserve the code space exactly.
Because $\{X_L,Z_L\}=0$ and $[B_L,B_R]=0$, the two
summands in $K$ anticommute.  Thus $U_c$ is a single
exponential and must \emph{not} be replaced by two sequential
contact exponentials.  The bath gates are nearest-neighbor,
whereas the code-preserving logical contact is an engineered
many-qubit physical operation.  \emph{No bath reset} occurs in
an individual, pure-state unitary trajectory.

%

\vspace{0.3cm}
\paragraph{Exact code-share decoupling and observables.}
For each code share $c_X$ containing at most two
physical qubits, the five-qubit erasure identity is
\begin{equation}
\operatorname{Tr}_{c_{\overline X}}
|\overline i\rangle\langle\overline j|
=\delta_{ij}\frac{\mathbb I_{c_X}}{d_{c_X}},
\qquad i,j\in\{0,1\}.
\label{eq:sm_bath_erasure}
\end{equation}
In the reduced logical--bath representation, any state can be written as
$|\Psi\rangle_{Lb}=\sum_{i=0}^1|i\rangle_L|\chi_i\rangle_b$.
Its physical encoding is
$|\Psi\rangle_{\mathrm{phys},b}
=(W\otimes\mathbb I_b)|\Psi\rangle_{Lb}
=\sum_{i=0}^1|\overline i\rangle|\chi_i\rangle_b$.
Applying Eq.~\eqref{eq:sm_bath_erasure} and tracing over $c_{\overline X} b_{\overline X}$ gives, for each individual
realization and at every time,
\begin{equation}
 \begin{aligned}
 \rho_X(t)=\frac{\mathbb I_{c_X}}{d_{c_X}}\otimes\rho_{b_X}(t),~~~~~~~
 S_n(X,t)=\ln d_{c_X}+S_n(\rho_{b_X}(t)).
 \end{aligned}
 \label{eq:sm_bath_fixed_share}
\end{equation}
In particular, $(d_{c_A},d_{c_B},d_{c_C})=(4,4,2)$.
The physical code shares remain maximally mixed and
uncorrelated with their respective bath shares, even
when the logical qubit is entangled with the entire bath.
The full-party marginals need not be stationary:
their remaining evolution comes from $\rho_{b_X}(t)$.
Therefore, this circuit has \emph{exact code-share protection},
not an exact fixed-spectrum plateau for the complete
parties $A,B,C$.

For each pure realization we calculate
$S_{1,\mathrm{av}}=\frac13\sum_X S_1(X)$,
the analogous $S_{2,\mathrm{av}}$, and the four-copy tripartite quantity $G\equiv GM_2^{(3)}$.
The nonlinear quantities are evaluated on each individual pure state \emph{before} averaging over random circuits.

\vspace{0.3cm}
\paragraph{Exact four-copy readout in the reduced logical--bath space.}
For the pure logical--bath state
$\varrho_{Lb}=|\Psi\rangle_{Lb}\langle\Psi|$, define
\begin{equation}
M_\mu=\operatorname{Tr}_{L b_C}
\left[(\sigma_\mu^{(L)}\otimes\mathbb I_b)\varrho_{Lb}\right],
\qquad \mu=0,x,y,z.
\label{eq:sm_bath_exact_readout_M}
\end{equation}
Each $M_\mu$ is Hermitian on $b_A b_B$. Set
$(R_\mu)_{aa',bb'}=(M_\mu)_{ab,a'b'}$.
Using Eq.~\eqref{eq:sm_typeII_Kcode} gives the exact physical-state
four-copy contraction
\begin{equation}
\begin{aligned}
Z_{2,\mathrm{tot}}^{(3)}
&=\operatorname{Tr}\!\left[
\varrho_{Lb}^{\otimes4}K_{\mathrm{code}}\mathcal P_b\right]\\
&=\frac1{64}\sum_{\mu=0,x,y,z}
\operatorname{Tr}\!\left[(R_\mu R_\mu^\dagger)^2\right],
\end{aligned}
\label{eq:sm_bath_exact_readout_Z}
\end{equation}
where $\mathcal P_b=P_{b_A}((13)(24))P_{b_B}((12)(34))$.
Writing $p_{b_X}=\operatorname{Tr}\rho_{b_X}^2$ and using the exact
code-share purity factors gives
\begin{equation}
G=\frac12\log\frac{p_{b_A}p_{b_B}p_{b_C}}
{32\,Z_{2,\mathrm{tot}}^{(3)}}.
\label{eq:sm_bath_exact_readout_G}
\end{equation}
These equations retain the joint logical--bath correlations of each
pure realization. They are used before ensemble averaging and do not
invoke the conditional Haar-isometry approximation.

\vspace{0.3cm}
\paragraph{Exactly averaged one-copy logical channel.}
The standard Haar-twirling identity for an
independently sampled two-qubit unitary $U_{ij}\sim\operatorname{Haar}[U(4)]$ gives, for an arbitrary incoming state $\rho_{Rij}$,
\begin{equation}
 \mathbb E_U[(\mathbb I_R\otimes U_{ij})\rho_{Rij}
 (\mathbb I_R\otimes U_{ij}^\dagger)]
 =\operatorname{Tr}_{ij}(\rho_{Rij})
 \otimes\frac{\mathbb I_{ij}}4.
 \label{eq:sm_bath_haar_twirl}
\end{equation}
Every bath site is acted on by at least one freshly sampled
gate in a complete even--odd period.  Successive conditional
averages therefore give the \emph{exact one-copy identity}
\begin{equation}
 \mathbb E_{U_b}\bigl[U_b\rho_{Lb}U_b^\dagger\bigr]
 =\rho_L\otimes\frac{\mathbb I_b}{2^9},
 \qquad \rho_L=\operatorname{Tr}_b\rho_{Lb}.
 \label{eq:sm_bath_onecopy_twirl}
\end{equation}
The incoming $\rho_{Lb}$ can be entangled and may depend
on all previous gates; only independence of the \emph{new}
gates is required.  Equation~\eqref{eq:sm_bath_onecopy_twirl}
does not describe a physical reset in any individual circuit.

Writing $c=\cos(\sqrt2\lambda)$ and
$s=\sin(\sqrt2\lambda)$, the identities
$B_L^2=B_R^2=\mathbb I$, $[B_L,B_R]=0$, and
$\{X_L,Z_L\}=0$ imply $K^2=2\mathbb I$ and hence
\begin{equation}
 U_c=c\mathbb I-i\frac{s}{\sqrt2}K.
 \label{eq:sm_bath_contact_exp}
\end{equation}
Because $\operatorname{Tr}_b B_L=
\operatorname{Tr}_b B_R=
\operatorname{Tr}_b(B_LB_R)=0$, tracing out the averaged
bath produces the exact logical channel
\begin{equation}
 \Phi_\lambda(\rho_L)=
 c^2\rho_L+\frac{s^2}{2}
 (X_L\rho_LX_L+Z_L\rho_LZ_L).
 \label{eq:sm_bath_logical_channel}
\end{equation}
In particular, $\Phi_\lambda(Z_L)=c^2Z_L$.
For $\rho_L(0)=(\mathbb I_L+Z_L)/2 = |0\rangle_L\langle 0|$, iteration yields
\begin{equation}
 \begin{split}
 \mathbb E\langle Z_L(t)\rangle
 =(c^2)^t=e^{-\Gamma_Lt},~~~~~~~~~~~
 \Gamma_L=-2\ln|\cos(\sqrt2\lambda)|
 =2\lambda^2+\frac23\lambda^4+O(\lambda^6).
 \end{split}
 \label{eq:sm_bath_Gamma}
\end{equation}
Thus $\tau_L=\Gamma_L^{-1}\sim(2\lambda^2)^{-1}$
is an independently derived slow timescale for the
\emph{ensemble-averaged logical polarization}.
This exact one-copy result does not determine the
nonlinear $G$: in general,
$\mathbb E[\rho^{\otimes4}]\neq (\mathbb E\rho)^{\otimes4}$.

\vspace{0.3cm}
\paragraph{Conditional four-copy Haar-isometry approximation.}
To estimate the contact-induced multipartite response,
we additionally consider an \emph{auxiliary} global
Haar-isometry ensemble.  After a fast bath transient,
condition on logical Schmidt weights
$p=(1+z)/2$ and $1-p$, and model the two orthonormal
bath Schmidt vectors as the first two columns of a Haar
unitary $U\in U(D)$, with $D=2^9=d^3$ and $d=8$:
\begin{equation}
 |\Psi(z,U)\rangle_{Lb}
 =\sqrt p\,|0\rangle_L U|0\rangle_b
 +\sqrt{1-p}\,|1\rangle_L U|1\rangle_b.
 \label{eq:sm_bath_haar_ansatz}
\end{equation}
This is a statistical closure for conditional states,
\emph{not} an assertion that the microscopic nearest-neighbor
circuit is globally Haar distributed.  We further identify
$z(t)=e^{-\Gamma_Lt}$ within this approximation.  The
latter is exactly the ensemble-mean $Z_L$ polarization,
but need not equal the Bloch-vector length in a single
realization.

With the replica permutation pattern defining
Eq.~\eqref{eq:prl-display-2}, the code kernel leads to
an auxiliary four-copy moment $\overline Z_4(z)$.
The Haar--Weingarten calculation yields
\begin{equation}
 \begin{split}
 \overline Z_4(z)
 =a_d(1+z^2)^2+b_d(1-z^2)^2,~~~~~~~
 a_d=\frac{3d^2(d^2+d+2)}
 {128(d^2-d+1)(d^3+2)(d^3+3)},~~~~~~~
 b_d=\frac{3}{128(d^4+d^2+1)}.
 \end{split}
 \label{eq:sm_bath_haar_Z4}
\end{equation}

\vspace{0.3cm}
\paragraph{General conditional Haar-isometry average and its error.}
To separate the exact conditional Haar calculation from its
microscopic-bath approximation, take bath dimensions
$a=b=c=8$, $D=abc=512$, and an arbitrary logical density matrix
$\rho_L=(\mathbb I+\boldsymbol r\cdot\boldsymbol\sigma_L)/2$.
Put $u=|\boldsymbol r|^2$ and $v=r_x^4+r_y^4+r_z^4$.
For a Haar random isometry
$V_{\mathrm H}:\mathbb C^2\to\mathbb C^{512}$,
let $|\psi_{\rho,V_{\mathrm H}}\rangle$ be the joint pure state obtained
by purifying $\rho_L$ through the two orthogonal bath columns of $V_{\mathrm H}$.
Then the logical marginal is exactly $\rho_L$; no mixed
logical Bloch vector is substituted into the pure-code identity.

For the code permutation kernel
$K_{\rm code}=64^{-1}\sum_\mu\sigma_\mu^{\otimes4}$ define
$A_\mu=\sqrt{\rho_L}\,\sigma_\mu\sqrt{\rho_L}$ and the cycle trace
$L_\tau=\sum_\mu\prod_{\gamma\in\mathrm{Cycles}(\tau)}
\operatorname{Tr}(A_\mu^{|\gamma|})$. With
$\alpha=(13)(24)$ and $\beta=(12)(34)$, the exact $S_4$
Weingarten average is
\begin{equation}
Z_{4,\mathrm{iso}}(\rho_L)=\frac1{64}
\sum_{\sigma,\tau\in S_4}\mathrm{Wg}_{D}(\sigma^{-1}\tau)
L_\tau\,a^{\#(\alpha^{-1}\sigma)}
 b^{\#(\beta^{-1}\sigma)}c^{\#\sigma}.
\label{eq:sm_bath_iso_general_sum}
\end{equation}
All $24^2=576$ signed permutation terms are included. For $D=512$ exact rational simplification gives 
\begin{equation}
Z_{4,\mathrm{iso}}=
\frac{305219+81018u-40509u^2+345728v}
{23497777280}.
\label{eq:sm_bath_iso_general_closed}
\end{equation}
The corresponding bath-share purity is
$p_X=8(69+4u)/4161$ for each of the three shares, and hence
$G_{\mathrm{iso}}=-\frac12\log Z_{4,\mathrm{iso}}
+\frac32\log p_X-\frac12\log32$.
For the special $z$-polarized conditional state
$u=z^2$, $v=z^4$, this exactly reduces to the previously
quoted $a_d,b_d$ formula and leaves its numerical plateau and
analytic crossing unchanged. This exact conditional average is
not by itself a theorem on the microscopic locally scrambling bath.

More precisely, let $M_k$ and $\widetilde M_k$ be the true
and auxiliary joint $k$-copy moments, and put
$\epsilon_k=\|M_k-\widetilde M_k\|_1$.
Since
$K_{\rm code}=(\mathbb I+X_L^{\otimes4})(\mathbb I+Z_L^{\otimes4})/64$
has operator norm $1/16$, and the bath permutation is unitary,
\begin{equation}
\begin{aligned}
|Z_4-\widetilde Z_4|&\le\epsilon_4/16,~~~~~~~
|P_{2,X}-\widetilde P_{2,X}|\le\epsilon_2.
\end{aligned}
\label{eq:sm_bath_conditional_momerr}
\end{equation}
If the corresponding relative errors
$\eta_4=\epsilon_4/(16\widetilde Z_4)$ and
$\eta_{2,X}=\epsilon_2/\widetilde P_{2,X}$ are less than one,
then the annealed logarithmic observables obey
\begin{equation}
|G_{\rm ann}-\widetilde G_{\rm ann}|
\le-\frac12\log(1-\eta_4)
-\frac12\sum_{X=A,B,C}\log(1-\eta_{2,X}).
\label{eq:sm_bath_conditional_bound}
\end{equation}
No bound making $\epsilon_4$ sufficiently small for the
actual local circuit has been established. Replacing each
trajectory's fluctuating Bloch vector by an exponentially decaying
mean-polarization direction is a separate closure approximation;
it must not be treated as exact four-copy dynamics.

The two-copy Haar twirl similarly gives the purity of a
single $d$-dimensional bath share,
\begin{equation}
 \overline P_b(z)=
 \frac{d(d^2+1)+\frac12d^2(1+z^2)}
 {d^4+d^2+1}.
 \label{eq:sm_bath_haar_purity}
\end{equation}
Including the exact code-share purity factors
$1/4,1/4,1/2$, define the auxiliary \emph{annealed} ratio
\begin{equation}
 G_H(z)=\frac12\ln\left[
 \frac{\overline P_b(z)^3}{32\,\overline Z_4(z)}
 \right].
 \label{eq:sm_bath_haar_G}
\end{equation}
For $d=8$ one has
$\eta=b_8/a_8=132355/172864$.
To model the paired $\lambda=0$ subtraction used in the numerical simulations, we subtract the uncoupled
Haar-isometry reference $G_H(1)$, for which the
logical qubit remains fully polarized ($z=1$)
while the bath is scrambled.
Defining $r=z^2=e^{-2\Gamma_L t}$, we obtain
the predicted contact-induced multi-entropy response
\begin{equation}
\Delta G_H(t)
=G_H(z(t))-G_H(1)=F(r)=\frac12\ln\!\left[
 \frac{4\bigl((69+4r)/73\bigr)^3}
 {(1+r)^2+\eta(1-r)^2}\right].
 \label{eq:sm_bath_F}
\end{equation}
In particular,
\begin{equation}
 \begin{split}
 F(1)&=0,\qquad F(0)=0.3243555790,\qquad
 F(r)=F(0)-0.04576457361\,r+O(r^2).
 \end{split}
 \label{eq:sm_bath_F_limit}
\end{equation}
The $r=e^{-2\Gamma_Lt}$ term is a late-time tail of this
\emph{auxiliary} four-copy closure, not a proven spectral
eigenvalue of the microscopic local-circuit four-copy
channel.  Moreover, Eq.~\eqref{eq:sm_bath_haar_G} places
logarithms \emph{after} auxiliary Haar averaging, whereas
the measured observable averages the logarithm computed
in each realization.  Replacing the latter by $F$ requires
both the conditional-frame approximation and sufficient
self-averaging of the paired nonlinear response.

The model's analytical $90\%$ crossing is defined relative
to its infinite-time value, $F(e^{-2\Gamma_L t_{90}^H})
=0.9F(0)$.  Solving the \emph{full} function gives
\begin{equation}
 t_{90}^H=\frac{0.7502141663}{\Gamma_L}
 \simeq0.3751071\,\lambda^{-2}\quad(\lambda\ll1).
 \label{eq:sm_bath_H_cross}
\end{equation}
The late-time expansion alone is not suitable for locating this crossing, since $\Gamma_Lt_{90}^H\simeq0.75$.

\vspace{0.3cm}
\paragraph{Paired finite-circuit simulations and operational clocks.}
We evolve $48$ independent bath-circuit realizations for each of
$\lambda=(0.025,0.032,0.042,0.055,0.072)$, together
with a $\lambda=0$ control.  Within each realization,
all six trajectories share the same initial bath product
state and the complete history of Haar gates.  Bath gates
are sampled using complex Gaussian matrices with QR
orthogonalization and diagonal-phase correction.
The respective maximum times are
\begin{equation}
 T_{\max}(\lambda)
 =8\left\lceil\frac{3.5}{8\Gamma_L(\lambda)}\right\rceil=(2800,1712,992,584,344).
 \label{eq:sm_bath_Tmax}
\end{equation}
The state advances through \emph{every} complete period;
observables are recorded at every integer time for
$0\le t\le20$, every four periods through $t=240$,
and every eight periods thereafter.  These recording
intervals do not alter the microscopic update.

For a trajectory $r$ and a coupling $\lambda$ define the
paired contact-induced response
\begin{equation}
 \begin{aligned}
 \Delta G_{r,\lambda}(t)&=G_{r,\lambda}(t)-G_{r,0}(t),~~~~~~~~
 \overline{\Delta G}_{\lambda}(t)=\frac1{48}\sum_{r=1}^{48}\Delta G_{r,\lambda}(t).
 \end{aligned}
 \label{eq:sm_bath_paired}
\end{equation}
The same averaging prescription applies to the total
$G_\lambda$ and to $S_{1,\mathrm{av}}$; importantly,
all nonlinear observables are calculated \emph{before}
the circuit average.  For each averaged observable
$\mathcal O(t)\in\{\overline{S}_{1,\mathrm{av}}(t),
\overline G_\lambda(t),\overline{\Delta G}_\lambda(t)\}$,
the reference $\mathcal O_{\mathrm{ref}}$ is its mean over
recorded points in the final scaled-time interval
$\Gamma_Lt\ge\Gamma_LT_{\max}-0.7$.
Its operational $90\%$ time is the earliest recorded
$t_j$ satisfying the threshold at three consecutive recorded points,
\begin{equation}
 t_{90}[\mathcal O]=\min\{t_j:\;
 \mathcal O(t_{j+k})\ge
 \mathcal O(0) +0.9[\mathcal O_{\mathrm{ref}}-\mathcal O(0)], k=0,1,2\}.
 \label{eq:sm_bath_t90}
\end{equation}
The total-$G$ and paired-$\Delta G$ crossing times are
computed separately, with their own measured references;
\emph{neither} uses $F(0)$ as its reference.

\begin{table}[h!]
 \centering
 \begingroup
\setlength{\tabcolsep}{10pt}
\renewcommand{\arraystretch}{1.3}
 \begin{tabular}{c|rrr|r}
 \hline\hline
 $\lambda$ & $t_{90}^{S}$ & $t_{90}^{G}$
 & $t_{90}^{\Delta G}$ & $t_{90}^{H}$ \\
 \hline
 $0.025$ & 5 & 360 & 616 & 600.0 \\
 $0.032$ & 5 & 224 & 368 & 366.2 \\
 $0.042$ & 5 & 128 & 224 & 212.5 \\
 $0.055$ & 5 & 80  & 124 & 123.9 \\
 $0.072$ & 5 & 44  & 72  & 72.2 \\
 \hline\hline
 \end{tabular}
 \endgroup
\caption{\small Measured operational $90\%$ crossing times
of the code-preserving local-Haar bath circuit.
All times are in complete even--odd--contact periods.
Here $t_{90}^{S}$ denotes the first sustained
$90\%$ growth crossing of the mean single-party
von Neumann entropy, $t_{90}^{G}$ refers to the total multi-entropy,
and $t_{90}^{\Delta G}$ to the paired contact-induced
multi-entropy response relative to $\lambda=0$.
The quantity $t_{90}^{H}$ is the independent,
parameter-free prediction from the auxiliary
four-copy Haar-isometry approximation.
The measured columns use 48 paired realizations; the last column is calculated from the auxiliary Haar-isometry approximation.}
 \label{tab:sm_bath_times} 
\end{table}

The mean ordinary von Neumann entropy reaches its
finite-window threshold after $5$ periods at all five
couplings, whereas the paired multipartite response
remains sensitive to the weak logical contact over a
much longer interval. Independent log--log regression
of the measured paired times gives
\begin{equation}
 t_{90}^{\Delta G}\simeq
 0.35189\,\lambda^{-2.02508},
 \qquad R_{\log t}^2=0.99939.
 \label{eq:sm_bath_fit}
\end{equation}
A 250-resample bootstrap over entire paired circuit
realizations gives a conditional $95\%$ interval
$[-2.072,-1.935]$ for the fitted exponent.
The measured late-window paired GM increments lie
between $0.32313$ and $0.32381$, close to the
independent auxiliary prediction $F(0)=0.32436$.
Using the exact $\Gamma_L(\lambda)$, with no free GM
amplitude or timescale, Eq.~\eqref{eq:sm_bath_H_cross}
predicts the last column of Table~\ref{tab:sm_bath_times}.
Across all couplings and recorded points at $t\ge12$,
the auxiliary function has descriptive agreement with
the paired mean trajectories ($R^2=0.99948$ and
RMSE $=0.00189$).  These are tests of an
approximation, not a proof of its assumptions.

The mean entropy has a small but nonzero late
contact-dependent correction; its fast $t_{90}^{S}$
therefore measures the \emph{dominant} bath-scrambling
growth, not the absence of all slower entropy response.
The exact code-share identity and the analytical
$\Gamma_L\propto\lambda^2$ establish, respectively,
protection of the encoded information and an independent
slow logical-polarization clock.  The observed
$t_{90}^{\Delta G}\propto\lambda^{-2}$ trend shows that
multipartite correlations couple strongly to that clock
within the studied finite-size circuit.  Because the bath
marginals still evolve, the complete Schmidt spectra are
not exactly frozen; neither the finite-window crossings
nor the auxiliary four-copy calculation establish ETH
thermalization or an exact asymptotic four-copy gap of
the microscopic local circuit.

\subsection{A logical memory distributed among three parties: exact collision circuit}
\label{sec:sm_typeII_collision}

\begin{figure*}[h!]
  \centering
  \includegraphics[width=0.68\textwidth]{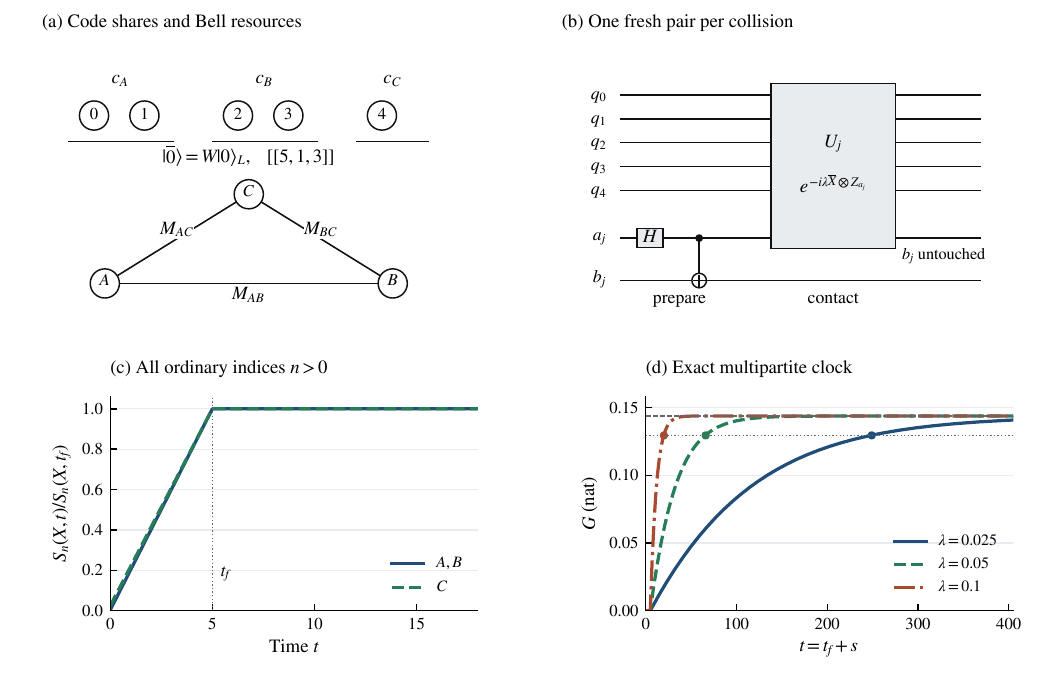}
  \caption{\small 
  Exact code-preserving collision circuit. A fast stage prepares Bell-pair resources; subsequent weak logical contacts preserve all three complete party density matrices while the four-copy invariant follows $G(s)=\frac12\log\!\left[\frac{4}{3+\cos^{2s}(4\lambda)}\right]$. The characteristic multipartite timescale scales as $\lambda^{-2}$, and the number of usable contacts is bounded by the finite fresh-pair supply.}
  \label{fig:qec_collision}
\end{figure*}

This is a repeated-interaction construction in the sense of quantum
collision models \cite{Ciccarello2022}. The ancillary Bell pairs remain
part of the complete pure state used to evaluate the multipartite
invariant. 
Figure~\ref{fig:qec_collision} shows the resource arrangement
and collision circuit in panels (a,b), and the ordinary-entropy
plateau and tunable multipartite dynamics in panels (c,d).

\vspace{0.3cm}
\paragraph{Code-space erasure identity.}
We use the same $[[5,1,3]]$ code and encoding isometry
$W:\mathcal H_L\to\mathcal H_{\mathrm{phys}}$ defined in
Sec.~\ref{sec:sm_typeII_static_code}. In contrast to the
physically perturbed static code, the collisions below act
strictly within the ideal code space. Partition the five
physical code qubits into $c_A=(0,1)$, $c_B=(2,3)$, and
$c_C=(4)$, and let $c_{\overline X}$ be the complement of
$c_X$ within the code. The correctable-erasure identity is
\begin{equation}
 \operatorname{Tr}_{c_{\overline X}}
 \bigl[|\overline i\rangle\langle\overline j|\bigr]
 =\operatorname{Tr}_{c_{\overline X}}
 \bigl[W|i\rangle_L\langle j|W^\dagger\bigr]
 =\delta_{ij}\frac{\mathbb I_{c_X}}{d_{c_X}},
 \label{eq:sm_typeII_erasure}
\end{equation}
where $i,j\in\{0,1\}$, $X\in\{A,B,C\}$, and
$(d_{c_A},d_{c_B},d_{c_C})=(4,4,2)$.
Consequently, even when the encoded logical qubit is
entangled with an arbitrary reference $R$, each code
share is decoupled from any reference subregion $R_X$:
$\rho_{c_X R_X}=(\mathbb I_{c_X}/d_{c_X})\otimes
\rho_{R_X}$. This is an exact code-space property.
\vspace{0.3cm}
\paragraph{Fast creation of ordinary entanglement.}
Supply $M_{AB}$, $M_{AC}$, and $M_{BC}$ ancillary
qubit pairs shared between the indicated parties.
Initially every ancillary pair is $|00\rangle$, while the physical
code qubits are in $|\overline0\rangle=W|0\rangle_L$.
A Hadamard gate on one half followed by an interparty
controlled-NOT prepares
$|\Phi^+\rangle=(|00\rangle+|11\rangle)/\sqrt2$.
Gates acting on disjoint pairs may be applied in parallel;
the preparation stage ends at $t_f$. If $k_{XY}(t)$
pairs shared between $X$ and $Y$ have been prepared
by time $t\le t_f$, the ordinary R\'enyi entropies are,
for every $n>0$,
\begin{equation}
 \begin{split}
 S_n(A,t)&=[2+k_{AB}(t)+k_{AC}(t)]\ln2,\\
 S_n(B,t)&=[2+k_{AB}(t)+k_{BC}(t)]\ln2,\\
 S_n(C,t)&=[1+k_{AC}(t)+k_{BC}(t)]\ln2.
 \end{split}
 \label{eq:sm_typeII_fast_entropies}
\end{equation}
The physical encoded state $|\overline0\rangle$ has
$G=0$, and an interparty Bell pair contributes no
genuine tripartite $G$ under the definition
in Eq.~\eqref{eq:prl-display-2}.  
Additivity thus gives
$G(t)=0$ throughout preparation. With
$M=M_{AB}+M_{AC}+M_{BC}$, the dimensions of the
\emph{complete} parties (code shares plus ancillas) are
\begin{equation}
d_A = 2^{2+M_{AB}+M_{AC}},\quad
 d_B = 2^{2+M_{AB}+M_{BC}},\quad
d_C = 2^{1+M_{AC}+M_{BC}},\quad
 d_A d_B d_C = 2^{5+2M}.
\label{eq:sm_typeII_dims}
\end{equation}
\vspace{0.3cm}
\paragraph{Slow, code-preserving collisions.}
After $t_f$, sequentially couple the logical qubit to
$s\le M_{AB}$ previously \emph{unused} $AB$ Bell pairs.
The $j$th physical contact acts on the code qubits and
the $A$ half $a_j$ of the selected Bell pair:
\begin{equation}
 U_j(\lambda)=\exp\!\left(-i\lambda\overline X
 \otimes Z_{a_j}\right) = \cos\lambda\, \mathbb I-i\sin\lambda\,\overline X\otimes Z_{a_j}.
 \label{eq:sm_typeII_collision_U}
\end{equation}
Here $\overline X$ acts on the five physical code qubits,
commutes with all stabilizers, and preserves the code space.
In the reduced logical representation the same contact is
$\widetilde U_j(\lambda)=\exp(-i\lambda X_L\otimes
Z_{a_j})$, since $W^\dagger\overline XW=X_L$.
Writing $\mathbf a=(a_1,\ldots,a_s)$ for the computational
bits of the \emph{used} $A$ halves, the exact \emph{physical}
encoded state on the code and the used Bell pairs is
\begin{equation}
 |\Psi_s\rangle_{\mathrm{phys}}
 =2^{-s/2}\sum_{\mathbf a\in\{0,1\}^s}
 \exp\!\left[-i\lambda\overline X
 \sum_{j=1}^s(-1)^{a_j}\right]
 \times|\overline0\rangle\otimes
 |\mathbf a\rangle_{A_u}|\mathbf a\rangle_{B_u},
 \label{eq:sm_typeII_state}
\end{equation}
with all untouched Bell pairs tensored on. The label $u$
denotes the used ancillas. Tracing over $B_u$ eliminates
terms with $\mathbf a\ne\mathbf a'$, so each $A_u$
computational string occurs with probability $2^{-s}$.
The same uniformity holds for $B_u$; untouched Bell pairs
also have maximally mixed individual halves. The erasure
identity, Eq.~\eqref{eq:sm_typeII_erasure}, then gives
\begin{equation}
 \rho_X(t_f+s)=\frac{\mathbb I_{d_X}}{d_X},
 \qquad X=A,B,C,\quad 0\le s\le M_{AB}.
 \label{eq:sm_typeII_exact_spectra}
\end{equation}
Thus \emph{all three complete reduced density matrices},
and not merely selected entropies, remain exactly fixed
throughout the collision stage.
\vspace{0.3cm}
\paragraph{Exact four-replica contraction.}
The code-sector replica contraction is determined
by the universal kernel in
Eq.~\eqref{eq:sm_typeII_Kcode},
with the code shares $c_A$ and $c_B$ corresponding
to the regions $A$ and $B$ of the isolated code.
The collision circuit additionally involves
replica contractions of the ancillary Bell pairs.

Inserting Eq.~\eqref{eq:sm_typeII_state}, a used Bell pair
yields the phase character
\begin{equation}
 \frac12\sum_{\xi=\pm1}e^{4i\lambda\xi}
 =\cos(4\lambda).
 \label{eq:sm_typeII_single_collision_character}
\end{equation}
For the used Bell pairs, introduce two independent uniform sign strings
$\boldsymbol\xi,\boldsymbol\eta\in\{\pm1\}^s$ and write
\begin{equation}
\theta_{\boldsymbol\xi}=\lambda\sum_{j=1}^s\xi_j,
\qquad
|v_{\boldsymbol\xi}\rangle
=e^{-i\theta_{\boldsymbol\xi}X_L}|0\rangle_L.
\end{equation}
The Bell-half permutations enforce the four ket strings to have the
pattern $(\boldsymbol\xi,\boldsymbol\eta,
\boldsymbol\eta,\boldsymbol\xi)$, with the bra strings interchanged.
Including the unused Bell pairs yields
\begin{equation}
Z_4(s)=\frac{4^{-M}}{64}\sum_{\mu=0,x,y,z}
\mathbb E_{\boldsymbol\xi,\boldsymbol\eta}
\left|\langle v_{\boldsymbol\eta}|\sigma_\mu^{(L)}
|v_{\boldsymbol\xi}\rangle\right|^4.
\label{eq:sm_collision_explicit_replicas}
\end{equation}
For $a=\theta_{\boldsymbol\xi}$ and $b=\theta_{\boldsymbol\eta}$,
\begin{equation}
\begin{aligned}
\sum_\mu|\langle v_b|\sigma_\mu^{(L)}|v_a\rangle|^4
&=\cos^4(a-b)+\sin^4(a-b)
+\cos^4(a+b)+\sin^4(a+b)=\frac32+\frac12\cos(4a)\cos(4b).
\end{aligned}
\end{equation}

Since
\[
\mathbb E_{\boldsymbol\xi}
\cos(4\theta_{\boldsymbol\xi})
=[\cos(4\lambda)]^s,
\]
independence gives the exact four-copy moment
\begin{equation}
Z_4(s)
=\frac{3+\cos^{2s}(4\lambda)}
       {128\,4^M}.
\label{eq:sm_typeII_Zexact}
\end{equation}
Using the fixed marginals and $d_Ad_Bd_C=32\,4^M$, we obtain
\begin{equation}
G(s)=-\frac12\ln\!\left[Z_4(s)d_Ad_Bd_C\right]
=\frac12\ln\frac{4}{3+\cos^{2s}(4\lambda)}.
\label{eq:sm_typeII_Gexact}
\end{equation}
This is an exact pure-state identity for any allowed
number of collisions $0\le s\le M_{AB}$, without noise
averaging or a Markov approximation.
\vspace{0.3cm}
\paragraph{Logical and multipartite rates; ancillary resources.}
Tracing out a used Bell pair defines a channel on the
\emph{abstract} logical density matrix $\rho_L$:
\begin{equation}
 \mathcal E_\lambda(\rho_L)
 =\cos^2\lambda\,\rho_L
 +\sin^2\lambda\,X_L\rho_LX_L.
 \label{eq:sm_typeII_logical_channel}
\end{equation}
For the initial abstract logical state $|0\rangle_L$
(physically encoded as $|\overline0\rangle$),
\begin{equation}
 \begin{split}
 &r_z(s)=\operatorname{Tr}[Z_L\rho_L(s)]
 =[\cos(2\lambda)]^s, \qquad
 \Gamma_L=-\ln|\cos(2\lambda)|
 =2\lambda^2+O(\lambda^4).
 \end{split}
 \label{eq:sm_typeII_rz}
\end{equation}
This logical-polarization decay rate is distinct from
the multipartite rate inferred from the four-copy result.
For $0<|\cos(4\lambda)|<1$ and a sufficient supply of fresh Bell pairs,
$G(s)$ approaches $G_\infty=\tfrac12\ln(4/3)$ with
\begin{equation}
 \begin{split}
 &\Gamma_G=-2\ln|\cos(4\lambda)|
 =16\lambda^2+O(\lambda^4),
 \qquad
 G_\infty-G(s)=\frac16e^{-\Gamma_Gs}
 +O(e^{-2\Gamma_Gs}).
 \end{split}
 \label{eq:sm_typeII_Grelax}
\end{equation}
Thus the preparation time $t_f$ can remain fixed while
the characteristic multipartite timescale
$\tau_G=\Gamma_G^{-1}=O(\lambda^{-2})$ grows.
For $t_f=5$ and $\lambda=0.05$, the first collision count
with $G(s)\ge0.9G_\infty$ is $s=61$, corresponding to
total time $t=66$. A finite circuit, however, permits at
most $s=M_{AB}$ such collisions: realizing a window of
order $\lambda^{-2}$ requires
$M_{AB}=O(\lambda^{-2})$ fresh Bell pairs. The assumed
$O(1)$ parallel Bell-pair preparation is not a claim of
constant-time preparation under arbitrary geometric
locality constraints. This engineered repeated-interaction
protocol does not establish ETH thermalization or an
asymptotic relaxation gap for a fixed local Hamiltonian.


\end{document}